\documentclass[11pt]{article}

\usepackage[dvips]{graphicx}
\usepackage{bm}
\usepackage{amsmath,amssymb}
\usepackage{comment}
\usepackage{cite}

\numberwithin{equation}{section}

\newcommand{\nt}{\nonumber\\}
\newcommand{\nn}{\nonumber}
\newcommand{\tx}{\text}

\newcommand{\brp}{{\bar p}}
\newcommand{\bmu}{{\bar\mu}}

\newcommand{\cN}{{\cal N}}

\newcommand{\cW}{{\cal W}}

\newcommand{\ha}{{\hat a}}
\newcommand{\hb}{{\hat b}}

\newcommand{\hj}{j^\circ}

\newcommand{\ba}{\begin{eqnarray}}
\newcommand{\ea}{\end{eqnarray}}
\newcommand{\sss}[1]{\subsubsection{#1}}
\newcommand{\sssn}[1]{\subsubsection*{#1}}
\newcommand{\back}{\!\!\!\!\!\!}

\newcommand{\alp}{\alpha}

\newcommand{\eps}{\epsilon}
\newcommand{\varkap}{\varkappa}
\newcommand{\Ups}{\Upsilon}
\newcommand{\ups}{\upsilon}

\newcommand{\emp}{\emptyset}

\newcommand{\valp}{{\vec\alpha}}
\newcommand{\vbet}{{\vec\beta}}
\newcommand{\vgam}{{\vec\gamma}}
\newcommand{\vrho}{{\vec\rho}}
\newcommand{\vome}{{\vec\omega}}
\newcommand{\vlam}{{\vec\lambda}}
\newcommand{\vphi}{{\vec\varphi}}
\newcommand{\va}{{\vec a}}
\newcommand{\vb}{{\vec b}}
\newcommand{\ve}{{\vec e}}

\newcommand{\vW}{{\vec W}}
\newcommand{\vY}{{\vec Y}}
\newcommand{\vha}{{\vec{\hat a}}}
\newcommand{\vhb}{{\vec{\hat b}}}

\newcommand{\bbC}{{\mathbb C}}
\newcommand{\bbZ}{{\mathbb Z}}

\newcommand{\cdotsb}{\cdots{\;\!\!}}
\newcommand{\parfrac}[2]{\frac{\partial{#1}}{\partial{#2}}}

\begin{document}

\begin{titlepage}
\begin{flushright}
KEK-TH-1506
\end{flushright}

\vskip 15mm

\begin{center}
{\bf\LARGE
Notes on 3-point functions of $A_{N-1}$ Toda theory\\[+12pt]
and AGT-W relation for $SU(N)$ quiver
}
\vskip 14mm
{\Large
Shotaro Shiba
}
\vskip 7mm
{\large\it
Institute of Particle and Nuclear Studies,\\[+2pt]
High Energy Accelerator Research Organization (KEK),\\[+2pt]
1-1 Oho, Tsukuba, Ibaraki 305-0801, Japan
}
\vskip 5mm
{\tt
sshiba@post.kek.jp
}
\end{center}

\vskip 28mm

\begin{abstract}
We study on the property of 3-point correlation functions of 2-dim $A_{N-1}$ Toda field theory, and show the correspondence with the 1-loop part of partition function of 4-dim $\cN=2$ $SU(N)$ quiver gauge theory.
As a result, we can check successfully the 1-loop part of AGT-W relation for all the cases of $SU(N)$ quiver gauge group. 
\end{abstract}

\end{titlepage}

\setcounter{footnote}{0}


\section{Introduction}

Recently, based on Gaiotto's discussion on $\cN=2$ dualities~\cite{Gaiotto:2009},
the relation between 4-dim $\cN=2$ gauge theories and the quantum geometry of 2-dim Riemann surface has become understood more clearly.
One of the most remarkable progress must be the proposition of AGT relation~\cite{Alday:2009}, which states that the partition function of 4-dim $\cN=2$ $SU(2)$ linear quiver gauge theory corresponds to the correlation function of 2-dim Liouville field theory.
As the natural generalization of this relation, the correspondence between $\cN=2$ $SU(N)$ linear quiver gauge theory and $A_{N-1}$ Toda field theory has been also proposed, which is called AGT-W relation~\cite{Wyllard:2009}.

The correspondence between the parameters of gauge theory and those of Toda theory in AGT-W relation has been already proposed for a general case of $SU(N)$ quiver~\cite{Kanno:2009, Drukker:2010}, but the proof is still incomplete.
Up to now, the proof by direct calculations has been done 
in the following cases:
For $SU(2)$ linear quivers, the correspondence has been checked for 
$SU(2)^n$ quiver with $n=1,2,3$ up to instanton level 3~\cite{Alba:2009}.
For $SU(3)$ linear quivers, it has been checked for $SU(3)^n$ quiver with $n=1,2$ and $SU(3)\times SU(2)$ quiver up to instanton level 3~\cite{Drukker:2010,Mironov:2009,Kanno:2010}.
For $SU(N)$ linear quivers with $N>3$, only the 1-loop part for $SU(N)\times$ $SU(N-1)\times \cdots \times SU(2)$ quiver has been discussed~\cite{Drukker:2010}.

On the conformal blocks in Toda theory which correspond to the instanton part of partition function in gauge theory, the discussion and calculation have been developed~\cite{Belavin:2011,Kanno:2011,Mironov:2010zs,Mironov:2010,Zhang:2011,Yanagida:2010,Itoyama:2010,Alday:2010,Kozcaz:2010,Tai:2010ps,Cheng:2010,Bonelli:2010,Mironov:2010a,Alba:2010,Piatek:2011,Mironov:2011,Shou:2011,Fateev:2011,Wyllard:2011,Fateev:2011a,Estienne:2011,Wu:2011}.
For example, by using the newly proposed basis with Young tableau indices~\cite{Belavin:2011} which is a kind of generalization of Jack polynomials,
we are getting to understand the reason why the factorized form of instanton partition function can be reproduced in Toda theory.
Now many researches restrict themselves to some limited cases, but they are very useful to deepen our understanding of the mechanism of AGT-W relation.

In this paper, on the other hand, we concentrate on the 1-loop part of partition function in gauge theory.
The corresponding part of correlation function in Toda theory is reduced to the product of 3-point correlation functions. 
In fact, we had the following problem in our previous paper~\cite{Kanno:2010}\:\!: 
when the two of three fields in a 3-point function are degenerate ones, some factors become zero and then make zeros and poles.
At that time, we could not find how to deal with, so we simply neglected them. 
In this paper, we reconsider carefully on this problem.
Then we grasp the mechanism of cancellations of undesirable factors including the zeros, and find the physical interpretation of the poles.
As a result, we can check successfully the 1-loop part of AGT-W relation for all the cases of $SU(N)$ quiver gauge group.

This paper is organized as follows.
In \S\,\ref{sec:notation} and \S\,\ref{sec:recur}, we review on the 3-point correlation function of $A_{N-1}$ Toda field theory. 
Then in \S\,\ref{sec:3pt-func}, we discuss the important properties of 3-point function for the proof of AGT-W relation.
In \S\,\ref{sec:check}, we check AGT-W relation by direct calculations:
We first summarize our ansatz in \S\,\ref{sec:ansatz}, then observe the correspondence in the cases of $A_2$ and $A_3$ Toda theory in \S\,\ref{sec:A2} and \S\,\ref{sec:A3}, and finally in \S\,\ref{sec:AN-1}, we discuss AGT-W relation for a general case of $A_{N-1}$ Toda theory in an algorithmic way.

\section{3-point correlation function of $A_{N-1}$ Toda theory}
\label{sec:Toda}

AGT-W relation is the nontrivial correspondence between
the partition function of 4-dim $\cN=2$ $SU(N)$ quiver gauge theory
and the correlation function of 2-dim $A_{N-1}$ Toda theory.
In general, we consider the quiver gauge theory with a chain
of $n$ $SU$ groups
\ba\label{quiver}
SU(d_1)\times SU(d_2)\times \cdots \times SU(d_{n-1})\times SU(d_n)\,.
\ea
Here we require that the theory should be conformal in the massless limit of matter fields by introducing additional hypermultiplets.
Since the number of these hypermultiplets must be non-negative, this requirement means that
\ba\label{quiver-cond}
k_a=(d_a-d_{a+1})-(d_{a-1}-d_a)\geq 0 \qquad \text{for}~{}^\forall a=1,\cdots,n-1\,,
\ea
then the ranks $d_a$ satisfy
\ba\label{group}
d_1\leq d_2\leq\cdots\leq d_{l-1}\leq d_l=\cdots=
d_r\geq d_{r+1}\geq\cdots\geq d_{n-1}\geq d_n\,.
\ea
For simplicity, in this paper, we concentrate on only the part of decreasing tail
\ba\label{group-cond}
N:=d_1\geq d_2\geq\cdots\geq d_{n-1}\geq d_n\,.
\ea
The generalization to the original case (\ref{group}) is almost straightforward.

In order to check AGT-W relation, 
we must calculate the partition function of gauge theory with quiver gauge group (\ref{quiver}), which is summarized in Appendix \ref{sec:Nek},
and the corresponding $(n+3)$-point correlation function of Toda theory:
\ba\label{diagram}
{\setlength{\unitlength}{1cm}
\begin{picture}(5,1.7)
 \put(0,0){\line(1,0){2.5}}
 \put(3.5,0){\line(1,0){2.5}}
 \put(1,0){\line(0,1){1}}
 \put(2,0){\line(0,1){1}}
 \put(4,0){\line(0,1){1}}
 \put(5,0){\line(0,1){1}}
 \put(-0.7,-0.1){\small$\vbet_\infty$}
 \put(0.8,1.2){\small$\vbet_{n+1}$}
 \put(1.8,1.2){\small$\vbet_1$}
 \put(3.8,1.2){\small$\vbet_{n-1}$}
 \put(4.8,1.2){\small$\vbet_n$}
 \put(6.2,-0.1){\small$\vbet_0$}
 \put(1.3,0.15){\small$\valp_1$}
 \put(4.3,0.15){\small$\valp_n$}
 \put(2.77,0.2){$\cdots$}
\end{picture}}
\ea
Here we believe that the momenta $\valp_j$, $\vbet_k$ of Toda vertex operators  correspond to the quiver gauge group (\ref{quiver})
by following our ansatz~\cite{Kanno:2009}.
The details will be reviewed in \S\,\ref{sec:ansatz}.
Then this correlation function can be calculated by pants decomposition,
{\em i.e.}~the decomposition into 3-point functions and propagators.
The 3-point functions are given in terms of $\Upsilon$-function, 
as we review in this section.
The propagators are given as the inverse Shapovalov matrices.
It is known that this matrix is infinite size, 
but is block diagonal with respect to each descendant level,
and each block is finite size.

In this paper, we concentrate on only the propagators with descendant level 0.
Then the correlation function (\ref{diagram}) is reduced to the product of $(n+1)$ 3-point functions, as we will see in eq.\,(\ref{tree}).
According to the proposition of AGT-W relation, this product of 3-point functions should correspond to the 1-loop part of partition function of gauge theory. In this section, before checking this correspondence, we briefly review the derivation of 3-point function and discuss its property.

\subsection{$A_{N-1}$ Toda theory}
\label{sec:notation}

First we summarize our notation for 2-dim $A_{N-1}$ Toda field theory.
The action is
\ba\label{S}
S=\int d^2\sigma\,\sqrt{g}\left[
\frac{1}{8\pi}g^{xy}\partial_x\vphi\cdot \partial_y\vphi
+\mu\sum_{k=1}^{N-1}e^{b\ve_k\cdot\vphi}
+\frac{Q}{4\pi}R\vrho\cdot\vphi
\right]
\ea
where $\vphi=(\varphi_1,\cdotsb,\varphi_N)$ is Toda field,
satisfying $\sum_{p=1}^N \varphi_p=0$.
$g_{xy}$ is the metric on 2-dim Riemann surface,
and $R$ is its curvature.
$\ve_k$ is the $k$-th simple root defined as
\ba\label{root}
\ve_k=(0,\cdotsb,0,1,-1,0,\cdotsb,0)
\ea
where $1$ is $k$-th element.
$\vrho$ is Weyl vector ({\em i.e.}~half the sum of all positive roots)
of $A_{N-1}$ algebra:
\ba
\vrho=\frac12(N-1,N-3,\cdotsb,3-N,1-N)\,.
\ea
$\mu$ is a scale parameter called the cosmological constant,
$b$ is a real parameter called the coupling constant, and $Q:=b+b^{-1}$.
The central charge of this conformal field theory is $c=(N-1)\left(1+N(N+1)Q^2\right)$.

The primary field, or the vertex operator, is defined as 
\ba
V_\valp(z):=e^{\valp\cdot\vphi(z)}
\ea
where $\valp=(\alp_1,\cdotsb,\alp_N)$ is called the momentum, 
satisfying $\sum_{p=1}^N \alpha_{p}=0$.
The 2-point correlation function (propagator) of vertex operators is normalized in the usual manner:
\ba
\langle
V_\valp(z_1)V_{2Q\vrho-\valp}(z_2)
\rangle
=\frac{1}{|z_{12}|^{4\Delta_\valp}}
\ea
where $\Delta_\valp$ is the conformal dimension of $V_{\valp}$,
and $z_{12}:=z_1-z_2$.
The 3-point correlation function 
must have standard coordinate dependence due to the conformal invariance:
\ba\label{3ptfunc}
\langle
V_{\valp_1}(z_1)V_{\valp_2}(z_2)V_{\valp_3}(z_3)
\rangle
=\frac{C(\valp_1,\valp_2,\valp_3)}{
 |z_{12}|^{2(\Delta_1+\Delta_2-\Delta_3)}
 |z_{23}|^{2(\Delta_2+\Delta_3-\Delta_1)}
 |z_{31}|^{2(\Delta_3+\Delta_1-\Delta_2)}
}
\ea
where $\Delta_i:=\Delta_{\valp_i}$.
It is known that the function $C(\valp_1,\valp_2,\valp_3)$ becomes perturbatively nonzero, only when the screening condition is satisfied~\cite{Fateev:2007}:
\ba\label{scr-cond}
\valp_1+\valp_2+\valp_3
+\sum_{k=1}^{N-1} bs_k \ve_k=2Q\vrho
\ea
where $s_k$ are non-negative integers, satisfying $s_1\leq s_2\leq \cdots\leq s_{N-1}$.
In this case, the function $C(\valp_1,\valp_2,\valp_3)$ has simple poles at each of the variables 
\ba\label{pole}
(2Q\vrho-\sum_{i=1}^3\valp_i,\vome_k)=bs_k 
\ea
where $\vome_k$'s are the dual basis to the simple roots, 
{\em i.e.}~$\ve_i\cdot\vome_j=\delta_{ij}$.
For example,
\ba
\vome_1=\frac{1}{N}(N-1,-1,\cdotsb,-1)\,,\quad
\vome_{N-1}=\frac{1}{N}(1,\cdotsb,1,1-N)\,.
\ea

\subsection{Derivation of 3-point function}
\label{sec:recur}

Now we briefly review the derivation of 3-point function.
As far as the author knows, the two ways of derivation have been discussed. 
One way is to use the recurrent relation for Coulomb integrals~\cite{Fateev:2007},
and the other way is to use the differential equation for 4-point functions~\cite{Fateev:2005}.

\sssn{Derivation 1\,: by recurrent relation for Coulomb integral}

We note that the main residue of the function $C(\valp_1,\valp_2,\valp_3)$ at the poles (\ref{pole}) can be written in terms of Coulomb integral $I$ as
\ba\label{C-I}
\overset{N-1}{\underset{k=1}\bigotimes}\,
\underset{(2Q\vrho-\sum\valp_i)\cdot\vome_k=bs_k}{\mathrm{res}}
C(\valp_1,\valp_2,\valp_3)
\!&=&\!
(-\pi\mu)^{s_1+\cdots+s_{N-1}}
\Bigl\langle  V_{\valp_1}(\infty)V_{\valp_2}(1)V_{\valp_3}(0)\prod_{k=1}^{N-1}\mathcal{Q}_k^{s_k}
\Bigr\rangle
\nt&=:&\!
(-\pi\mu)^{s_1+\cdots+s_{N-1}}\,
I_{s_1\cdots\,s_{N-1}}(\valp_1,\valp_2,\valp_3)
\ea
where $\mathcal{Q}_k=\int d^2z\,e^{b \ve_k\cdot\vphi}$ is a screening charge.
As we will see in \S\,\ref{sec:ansatz}, for the check of AGT-W relation, we need only to consider the particular cases of 3-point function where one of the momenta $\valp_3=\varkappa\vome_1$ or $\varkappa\vome_{N-1}$
($\varkappa\in \bbC$).
These two momenta are related by the conjugation, and here we choose $\valp_3=\varkappa\vome_{N-1}$. Then the Coulomb integral in (\ref{C-I}) becomes
\ba\label{J}
&&\back
I_{s_1\cdots\,s_{N-1}}(\valp_1,\valp_2,\varkappa\vome_{N-1})
\\&&
=\int \prod_{k=1}^{N-1} d\mu_{s_k}(t_k)\,
 \mathcal{D}_{s_k}^{-2b^2}(t_k)
 \prod_{j=1}^{s_k}\,
 \bigl|t^{(j)}_{N-1}\bigr|^{-2b\varkap}\,
 \bigl|t^{(j)}_k-1\bigr|^{-2b\valp_2\cdot\ve_k}
 \prod_{l=1}^{N-2}\prod_{i=1}^{s_l}\prod_{i'=1}^{s_{l+1}}\,\bigl|t^{(i)}_l-t^{(i')}_{l+1}\bigr|^{2b^2}
\nn
\ea
where $t_k^{(j)}$ is the coordinate of the $j$-th screening field $e^{b \ve_k\cdot\vphi}$,
and we define
\ba
&&\back
d\mu_{s_k}(t_k)=\frac{1}{\pi^{s_k}s_k!}\prod_{j=1}^{s_k}d^2t_k^{(j)}\,,
\quad
\mathcal{D}_{s_k}(t_k)=\prod_{i<j}^{s_k}\,\bigl|t_k^{(i)}-t_k^{(j)}\bigr|^2\,.
\ea
By using the identity 
\ba\label{identity-D}
&&\back\!\!\!
\int d\mu_{s_k}(t_k)\,\mathcal{D}_{s_k}(t_k)\prod_{i=1}^{s_k}\prod_{j=1}^{s_k+s_l+1}\!
\big| t_k^{(i)}-x^{(j)}\big|^{2p_j}
\\&&\!\!\!\!\!\!\!\!\!=
\frac{\prod_j \gamma(1+p_j)} 
{\gamma(1+s_k+\sum_j p_j)}
\prod_{j'<j}^{s_k+s_l+1}\!\big| x^{(j')}-x^{(j)}\big|^{2+2p_{j'}+2p_j}
\int d\mu_{s_l}(t_l)\,\mathcal{D}_{s_l}(t_l)\prod_{i'=1}^{s_l}\prod_{j=1}^{s_k+s_l+1}\!
\big| t_l^{(i')}-x^{(j)}\big|^{-2p_j-2}\nn
\ea
where 
$\gamma(x)=\Gamma(x)/\Gamma(1-x)$,
we can derive the recurrent relation for the integral (\ref{J}) as
\ba\label{recur1}
I_{s_1\cdots\,s_{N-1}}(\valp_1,\valp_2,\varkap\vome_{N-1})
=
K(\valp_2,\varkap)\,
I_{s_1-1,\cdotsb,s_{N-1}-1}(\valp_1',\valp_2+b\vome_1,(\varkap+b)\vome_{N-1})
\ea
where $\valp_1'$ is determined by the screening condition (\ref{scr-cond}) and
\ba\label{K}
K(\valp_2,\varkap)
\!&=&\!
 \frac{\gamma(-s_1b^2)}{\gamma^{N-1}(-b^2)}
 \frac{\gamma(1-b\varkap)\,\gamma(N-1-b\sum_{k=1}^{N-1}\valp_2\cdot\ve_k+(N-2)b^2)}
 {\gamma(N-b\varkap-b\sum_{k=1}^{N-1}\valp_2\cdot\ve_k+(N-1-s_{N-1})b^2)}
\nt&&\!\times
 \prod_{j=1}^{N-2}
 \frac{\gamma(j-b\sum_{k=1}^{j}\valp_2\cdot\ve_k+(j-1)b^2)}
 {\gamma(1+j-b\sum_{k=1}^{j}\valp_2\cdot\ve_k+(s_{j+1}-s_{j}+j)b^2)}\,.
\ea
From this recurrent relation,
we can obtain the expression of Coulomb integral as
\ba
I_{s_1\cdots s_{N-1}}(\valp_1,\valp_2,\varkap\vome_{N-1})
=\left[\frac{-1}{\gamma(-b^2)}\right]^{s_1+\cdots+s_{N-1}}
\prod_{q=0}^{s_{N-1}}\frac{1}{\gamma(b\varkap+qb^2)}
\,
R^{s_1}_{1}
\prod_{j=2}^{N-1} R^{s_{j}-s_{j-1}}_{j}
\ea
where
\ba
R^{s}_{k}:=\prod_{p=1}^s \gamma(-pb^2)
\prod_{i=1,2}\,
\prod_{j\geq k}^N \gamma\left((Q\vrho-\valp_i)\cdot(\vlam_j-\vlam_k)b-pb^2\right)
;\quad
\vlam_k:=\vome_1-\sum_{\ell=1}^{k-1}\ve_\ell\,.
\ea
Here $\vlam_k$ are the fundamental weights of $sl(N)$ Lie algebra.
The invariant terms under the recurrent relation (\ref{recur1}) are determined by the symmetry under the exchange of $\valp_1\leftrightarrow \valp_2$.

Up to now, we neglect the dual screening charges $\tilde{\mathcal Q}_k=\int d^2z\,e^{b^{-1}\ve_k\cdot\vphi}$.
It has no problem for the classical arguments, but in quantum theory, we must consider them: the term
$\tilde \mu\sum_{k=1}^{N-1} e^{b^{-1}\ve_k\cdot\vphi}$
 in the action (\ref{S}) must be added, and $bs_k$ in eq.\,(\ref{scr-cond}) and (\ref{pole}) must be modified as $bs_k+b^{-1}\tilde s_k$.
Here
$
\tilde \mu:=\left(\pi\mu\gamma(b^2)\right)^{1/b^2}\!\!{\big/}{\pi\gamma(b^{-2})}
$
is the dual cosmological constant, and $s_k\, (\tilde s_k)$ is the number of (dual) screening charges.
Especially, this generalization means that the 3-point function must be invariant under the transformation $b\leftrightarrow b^{-1}$.
Then, as an entire self-dual function with respect to this transformation, we should consider the $\Upsilon$-function. Its integral representation is given as
\ba
\log\Upsilon(x)=\int_0^\infty \frac{dt}{t}\left[\left(\frac{Q}{2}-x\right)^2e^{-2t}-\frac{\sinh^2\left(Q/2-x\right)t}{\sinh bt \cdot \sinh b^{-1}t}\right]
\ea
which is convergent only in the strip $0<\mathrm{Re}\;x<Q$, otherwise we take an analytic continuation.
This $\Upsilon$-function is related to $\gamma$-function through the recurrent relation 
\ba
\Upsilon(x+b)=\gamma(bx)b^{1-2bx}\Upsilon(x)\,,\quad
\Upsilon(x+b^{-1})=\gamma(x/b)b^{2x/b-1}\Upsilon(x)\,,
\ea
where the normalization condition is $\Upsilon(Q/2)=1$.
Then by using
eq.\,(\ref{C-I}),
the formula of 3-point function can be finally proposed as~\cite{Fateev:2007}
\ba\label{3pt}
C(\valp_1,\valp_2,\varkap \vome_{N-1})
\!&=&\!
\left[\pi\mu\gamma(b^2)b^{2-2b^2}\right]^{(2Q\vrho-\valp_1-\valp_2-\varkappa\vome_{N-1})\cdot\vrho/b}
\nt&&\times
\frac{\Upsilon(b)^{N-1}\Upsilon(\varkap)\prod_{e>0}\Upsilon\left((Q\vrho-\valp_1)\cdot\ve\right)\Upsilon\left((Q\vrho-\valp_2)\cdot\ve\right)}{\prod_{j,k}\Upsilon\left(\varkap/N+(\valp_1-Q\vrho)\cdot\vlam_j+(\valp_2-Q\vrho)\cdot\vlam_k\right)}
\ea
where $e>0$ means all the positive roots.

Now we have some comments from the detail of calculation.
We note that this discussion holds good, even after the analytic continuation 
to the non-integer values $s_k$.
Such analytic continuation is often taken in the expression for the 3-point functions, as we will see in \S\,\ref{sec:3pt-func}. 
We also note that the recurrent relation (\ref{recur1}) can be satisfied,
only when 
\ba\label{recur-cond}
(Q\vrho-\valp_2)\cdot\sum_{l=1}^k \ve_l\neq Q
\qquad\text{for}~{}^\forall k=1,\cdotsb,N-1\,.
\ea
Then in the cases where this condition is violated,
the formula (\ref{3pt}) needs to be justified by another way of derivation.

\sssn{Derivation 2\,: by differential equation for 4-point function}

The other way of derivation is to use the differential equation~\cite{Fateev:2005}.
It is known that a certain kind of 4-point function of $A_{N-1}$ Toda field theory can be written as
\ba\label{4pt}
\left\langle V_{-b\vome_1}(x)V_{\valp_1}(0)V_{\valp_2}(\infty)V_{\varkappa\vome_{N-1}}(1)
\right\rangle
=|x|^{2b\valp_1\cdot\vlam_1}|1-x|^{2b\varkappa/N}G(x,\bar x)
\ea
where no conditions are imposed on momenta $\valp_1$, $\valp_2$ and
a complex number $\varkap$.
The function $G(x)$ satisfies the generalized Pochhammer hypergeometric equation
\ba\label{dif-eq}
\left[x\cdot\prod_{i=1}^N \left(x\parfrac{}{x}+A_i\right)
 -\prod_{i=1}^N\left(x\parfrac{}{x}+B_i-1\right)\cdot x\parfrac{}{x}\right]G(x,\bar x)=0
\ea
where
\ba
A_k\!&=&\!\frac{b\varkappa}{N}-\frac{N-1}{N}b^2
+b(\valp_1-Q\vrho)\cdot\vlam_1+b(\valp_2-Q\vrho)\cdot\vlam_k
\nt
B_k\!&=&\!1+b(\valp_1-Q\vrho)\cdot(\vlam_1-\vlam_{k+1})\,.
\ea
Since $G(x,\bar x)$ should satisfy the same equation (\ref{dif-eq}) with the replacement of $x\to \bar x$ (the complex conjugation of $x$), we can obtain the integral representation as
\ba\label{int-rep}
G(x,\bar x)=\int\prod_{i=1}^{N-1}d^2t_i\,
|t_i|^{2(A_i-B_i)}|t_i-t_{i+1}|^{2(B_i-A_{i+1}-1)}|t_1-x|^{-2A_1}
\ea
up to an overall constant. Here we set $t_N=1$.

In order to obtain the formula of 3-point function which we are interested in,
we take a 
decomposition for the 4-point function (\ref{4pt})
by using OPE
\ba
V_{-b\vome_1}(x)V_{\valp_k}(0)
=\sum_{j=1}^N C^{\valp_k-b\vlam_j}_{-b\vome_1,\valp_k}
\left(|x|^{2\Delta_{kj}}V_{\valp_k-b\vlam_j}(0)+\cdots\right)
\ea
where $\Delta_{kj}:=\Delta_{\valp_k-b\vlam_j}-\Delta_{-b\vome_1}-\Delta_{\valp_k}$,
and `$\cdotsb$' includes the contribution of descendant fields.
Then we can rewrite eq.\,(\ref{4pt}) as
\ba\label{4pt-2}
&&\back
\langle V_{-b\vome_1}(x)V_{\valp_1}(0)V_{\valp_2}(\infty)V_{\varkappa\vome_{N-1}}(1)
\rangle
\nt&&
=|x|^{2b\valp_1\cdot\vlam_1}|1-x|^{2b\varkappa/N}
\sum_{j=1}^N C^{\valp_1-b\vlam_j}_{-b\vome_1,\valp_1}C(\valp_1-b\vlam_j,\valp_2,\varkappa\vome_{N-1})|G_j(x)|^2
\ea
where $C(\valp_1,\valp_2,\valp_3)$ is a 3-point function defined in eq.\,(\ref{3ptfunc}), and $G_j(x)$ can be expressed in terms of the generalized hypergeometric function of type $(N,N-1)$ as
\ba
G_1(x)\!&=&\!
F \left(
\begin{array}{c|c}
\begin{matrix}
\!\! A_1\cdotsb A_N \\
\!\! B_1\cdotsb B_{N-1}
\end{matrix}
& x \!\!
\end{array}\right)
\nt
G_{k+1}(x)\!&=&\!
x^{1-B_k}\,
F \left(
\begin{array}{c|c}
\begin{matrix}
\!\! 1+A_1-B_k,\cdotsb,1+A_N-B_k\\
\!\! 1+B_1-B_k,\cdotsb,2-B_k,\cdotsb,1+B_{N-1}-B_k
\end{matrix}
& x \!\!
\end{array}\right)
\ea
for $k=1,\cdotsb,N-1$, and where
\ba
F\left(
\begin{array}{c|c}
\begin{matrix}
\!\! A_1\cdotsb A_N \\
\!\! B_1\cdotsb B_{N-1}
\end{matrix}
& x \!\!
\end{array}\right)
=1+\frac{\prod_{j=1}^N A_j}{\prod_{k=1}^{N-1} B_k}x
  +\frac{\prod_{j=1}^N A_j(A_j+1)}{\prod_{k=1}^{N-1} B_k(B_k+1)}\frac{x^2}{2}
 +\cdots\,.
\ea
Therefore, by comparing eq.\,(\ref{4pt-2}) with eq.\,(\ref{4pt}) using the integral representation (\ref{int-rep}), we can find the relation
\ba
\frac{C^{\valp_1-b\vlam_1}_{-b\vome_1,\valp_1}C(\valp_1-b\vlam_1,\valp_2,\varkappa\vome_{N-1})}
{C^{\valp_1-b\vlam_k}_{-b\vome_1,\valp_1}C(\valp_1-b\vlam_k,\valp_2,\varkappa\vome_{N-1})}
=\prod_{j=1}^N \frac{\gamma(A_j)\gamma(B_{k-1}-A_j)}{\gamma(B_j)\gamma(B_{k-1}-B_j)}
\ea
where we set $B_0=B_N=1$.
The structure constants $C^{\valp_1-b \vlam_k}_{-b\vome_1,\valp_1}$
can be calculated explicitly by the free field representation~\cite{Fateev:2000}
\ba\label{C-OPE}
C^{\valp_1-b \vlam_k}_{-b\vome_1,\valp_1}
=\left[-\frac{\pi\mu}{\gamma(-b^2)}\right]^{k-1}\,
 \prod_{i=1}^{k-1}\frac{\gamma(b(\valp_1-Q\vrho)\cdot(\vlam_i-\vlam_k))}{\gamma(1+b^2+b(\valp_1-Q\vrho)\cdot(\vlam_i-\vlam_k))}\,,
\ea
so we can properly obtain the ratio
$C(\valp_1-b\vlam_k,\valp_2,\varkappa\vome_{N-1})/C(\valp_1-b\vlam_1,\valp_2,\varkappa\vome_{N-1})$.
By using this relation, and 
by a similar discussion of the dual screening charge in \S\,\ref{sec:recur},
we can finally reproduce the formula (\ref{3pt}).

This way of derivation is rather complicated, but is superior in that it can justify the formula including the cases where the condition (\ref{recur-cond}) in the previous derivation is violated.
In fact, this condition means only that the denominator of eq.\,(\ref{C-OPE}) diverges in the present derivation, 
which has no problem for calculating the 3-point function.

\sssn{Results}

The formula of 3-point function with one of the momenta
$\valp_3=\varkap\vome_{N-1}$ is 
\ba\label{3pt-s}
C(\valp_1,\valp_2;\varkap \vome_{N-1})
\!&=&\!
\left[\pi\mu\gamma(b^2)b^{2-2b^2}\right]^{(2Q\vrho-\valp_1-\valp_2-\varkappa\vome_{N-1})\cdot\vrho/b}
\nt&&\times
\frac{\Upsilon(b)^{N-1}\Upsilon(\varkap)\prod_{e>0}\Upsilon((Q\vrho-\valp_1)\cdot\ve)\Upsilon((Q\vrho-\valp_2)\cdot\ve)}{\prod_{i,j}\Upsilon\left(\varkap/N+(\valp_1-Q\vrho)\cdot\vlam_i+(\valp_2-Q\vrho)\cdot\vlam_j\right)}\,.
\ea
The formula with one of the momenta $\valp_3=\varkap\vome_1$ can be obtained by taking the conjugation $\vlam_k\to{\vlam_k}^*=-\vlam_{N+1-k}$
(then $\vome_{N-1}^*=\vome_1$) as
\ba
C(\valp_1,\valp_2;\varkap \vome_1)
\!&=&\!
\left[\pi\mu\gamma(b^2)b^{2-2b^2}\right]^{(2Q\vrho-\valp_1-\valp_2-\varkappa\vome_1)\cdot\vrho/b}
\nt&&\times
\frac{\Upsilon(b)^{N-1}\Upsilon(\varkap)\prod_{e>0}\Upsilon((Q\vrho-\valp_1)\cdot\ve)\Upsilon((Q\vrho-\valp_2)\cdot\ve)}{\prod_{i,j}\Upsilon\left(\varkap/N-(\valp_1-Q\vrho)\cdot\vlam_i-(\valp_2-Q\vrho)\cdot\vlam_j\right)}\,.
\ea
Here $e>0$ means all the positive roots, and $i,j=1,\cdotsb,N$.

\subsection{Property of 3-point function}
\label{sec:3pt-func}

Before beginning the proof of AGT-W relation, we observe and discuss some properties of 3-point function.
As we mentioned in \S\,\ref{sec:notation}, the function $C(\valp_1,\valp_2,\valp_3)$ has simple poles where eq.\,(\ref{scr-cond}) is satisfied. 
In fact, the denominator of 3-point function (\ref{3pt-s}) can be written as
\ba\label{zeros-0}
\prod_{j_1,\,j_2=1}^{N}
\Upsilon\left((\valp_1-Q\vrho)\cdot\vlam_{j_1}+(\valp_2-Q\vrho)\cdot\vlam_{j_2}+\varkap\vome_{N-1}\cdot\vlam_{j_3}\right)
\ea
for ${}^\forall j_3=1,\cdotsb,N-1$,
and the $\Upsilon$-function has zeros at
\ba
\Upsilon(x)=0\quad\Leftrightarrow\quad
x=-mb-nb^{-1}\,,~ (m+1)b+(n+1)b^{-1}
\quad
\left(m,n\in \bbZ_{\geq 0}\right)\,.
\ea
Then we can find that 
all the factors with $j_1=j_2=j_3=:j$ in eq.\,(\ref{zeros-0}) become
\ba\label{zeros}
\Upsilon\left(
b(s_{j-1}-s_j)+b^{-1}(\tilde s_{j-1}-\tilde s_j)
\right)=0
\qquad\text{for}~{}^\forall j=1,\cdotsb,N-1\,,
\ea
when the screening condition (\ref{scr-cond}) is satisfied.
Here $s_k$, $\tilde s_k$ are integers, and we set $s_0=\tilde s_0=0$.
Therefore, in this case, the whole denominator (\ref{zeros-0}) can be rewritten as
\ba\label{extra-zeros}
&&\back
\prod_{j=1}^{N-1}
\Upsilon\left(
b(s_{j-1}-s_j)+b^{-1}(\tilde s_{j-1}-\tilde s_j)\right)
\times \Upsilon(\varkappa)
\nt&&\back\,\,\,\times
\prod_{j_1>j_2'}
\Upsilon\left((\valp_1-Q\vrho)\cdot(\vlam_{j_1}-\vlam_{j_2'})\right)
\prod_{j_2>j_1'}
\Upsilon\left((\valp_2-Q\vrho)\cdot(\vlam_{j_2}-\vlam_{j_1'})\right)
\ea
where $j_1,j_2=1,\cdotsb,N$ and $j_1',j_2'=1,\cdotsb,N-1$.
Now we can easily find that all the factors in the second line are canceled by the factors 
\mbox{$\Upsilon((Q\vrho-\valp_i)\cdot\ve)$} in the numerator,
since $\vlam_{j'}-\vlam_j=-\sum_{k=j}^{j'-1} \ve_k$ for $j'>j$.
As a result, when the screening condition (\ref{scr-cond}) is satisfied,
the 3-point function (\ref{3pt-s}) can be simplified as
\ba
C(\valp_1,\valp_2;\varkap\vome_{N-1})
\,=\,
\left[\pi\mu\gamma(b^2)b^{2-2b^2}\right]^{(2Q\vrho-\valp_1-\valp_2-\varkappa\vome_{N-1})\cdot\vrho/b}
\left[\frac{\Upsilon(b)}{\Upsilon(0)}\right]^{N-1}\,.
\ea

However, the story becomes a little more complicated in AGT-W relation.
Let us now discuss it, since it is very important for the proof of AGT-W relation.

\sssn{Discussion for the proof of AGT-W relation}

In AGT-W relation, we usually take an analytic continuation of the number of screening charges $s_k$, $\tilde s_k$ to non-integer values.
The formula of 3-point function (\ref{3pt-s}) is still valid in this case,
as we mentioned.
However, the number and the positions of poles must be changed, 
since the factors (\ref{zeros}) generally don't become zero.

If there are no zeros in the denominator, 
the zeros in the numerator mean that the 3-point function vanishes.
In fact, as we will see in \S\,\ref{sec:check}. 
such a situation is realized in the case of $SU(N)^n$ quiver 
({\em i.e.}~without descending nor ascending tails).
The vanishing 3-point function must be inadequate to compare with the partition function of gauge theory.
Therefore, we must avoid such a setting of parameters
by imposing the condition 
\ba\label{cond-notail}
\Upsilon((Q\vrho-\valp_i)\cdot\ve)\neq 0
\qquad\text{for}~{}^\forall i=1,2\,,~{}^\forall e>0\,.
\ea


For the quivers with descending or ascending tails,
on the other hand,
we always set new poles by imposing the following condition on some factors in the denominator (\ref{zeros-0}):
\ba\label{zeros-1}
\Upsilon\left(\upsilon_{\hj_1,\hj_2}\right)
:=\Upsilon\left((\valp_1-Q\vrho)\cdot\vlam_{\hj_1}+(\valp_2-Q\vrho)\cdot\vlam_{\hj_2}+\varkap\vome_{N-1}\cdot\vlam_{\hj_3}\right)=0
\ea
for ${}^\exists \hj_1,\hj_2=1,\cdotsb,N$ and ${}^\forall \hj_3=1,\cdotsb,N-1$.
Here we don't require $\hj_1=\hj_2=\hj_3$, which is different from the discussion before an analytic continuation.
In fact, this condition (\ref{zeros-1}) is indispensable to match with the partition function of gauge theory.

\begin{figure}[t]
\begin{center}
{\setlength{\unitlength}{1mm}
\begin{picture}(100,35)
\put(0,12){\line(1,0){70}}
\put(0,15){\line(1,0){100}}
\put(0,18){\line(1,0){100}}
\put(0,21){\line(1,0){100}}
\put(25,-1){\line(0,1){35}}
\put(39,-1){\line(0,1){35}}
\put(61,-1){\line(0,1){35}}
\put(75,-1){\line(0,1){6.2}}
\put(75,8){\line(0,1){26}}
\put(-1.6,11.1){$\bm\otimes$}
\put(-1.6,14.1){$\bm\otimes$}
\put(-1.6,17.1){$\bm\otimes$}
\put(-1.6,20.1){$\bm\otimes$}
\put(68.4,11.1){$\bm\otimes$}
\put(98.4,14.1){$\bm\otimes$}
\put(98.4,17.1){$\bm\otimes$}
\put(98.4,20.1){$\bm\otimes$}
\put(47.5,27){$\cdots$}
\put(47.5,4){$\cdots$}
\put(-25,0){\vector(1,0){7}}
\put(-25,0){\vector(0,1){7}}
\put(-26,8.5){$x^{4,5}$}
\put(-16.5,-0.7){$x^6$}
\put(93,9){\vector(-1,0){24}}
\put(71.5,5.5){\footnotesize{HW\,transition}}
\end{picture}}
\caption{Brane configuration of D4-\,(horizontal), NS5-\,(vertical), D6-\,($\bm\otimes$)\,branes}
\label{fig:brane}
\end{center}
\end{figure}
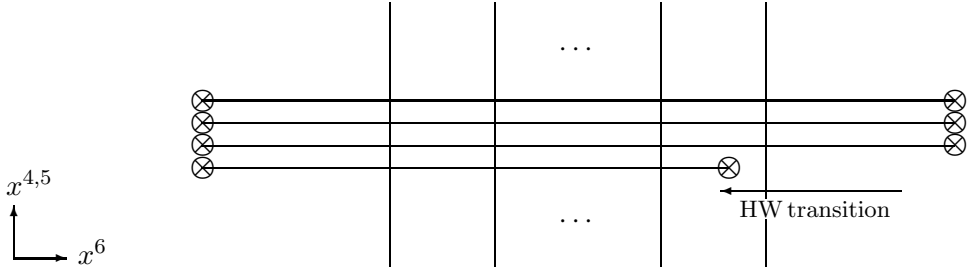

In order to discuss the physical meaning of the condition (\ref{zeros-1}), it seems convenient to consider the brane configuration~\cite{Witten:1997}.
For the quivers without tails, we consider the system of intersecting D4-branes and NS5-branes.
For the quivers with tails, on the other hand, D6-branes are introduced from the infinite distance in the D4/NS5 system.
The directions where each kind of branes is extended are shown in the following table:
\begin{center}
\vspace{2mm}
\begin{tabular}{|c||c|c|c|c|}
\hline
&$0,1,2,3$&$4,5$&$6$&$7,8,9$\\\hline
D4-branes&$-\!\!-$&&$-\!\!-$&\\\hline
NS5-branes&$-\!\!-$&$-\!\!-$&&\\\hline
D6-branes&$-\!\!-$&&&$-\!\!-$\\\hline
\end{tabular}
\vspace{2.5mm}
\end{center}
The arrangement of branes is shown in figure~\;\!\!\ref{fig:brane}.
Here we consider the situation where each D6-brane in the infinite distance $|x^6|=\infty$ is coupled to an infinitely extended D4-brane.
Then the D6-branes are moved to $|x^6|=$ finite, passing through some of the NS5-branes.
At this time, they cause Hanany-Witten transition~\cite{Hanany:1996} on D4-branes, which makes the number of D4-branes ({\em i.e.}~the rank of gauge group) smaller.
Moreover, superstrings ending on a D4-brane and a D6-brane (4-6 strings) appear, and they behave as new fundamental (or antifundamental) matter fields in 4-dim gauge theory defined in the $x^{0,1,2,3}$ spacetime.
The mass of these new fields is determined by the position of D6-branes.
Here we note that the condition (\ref{zeros-1}) in Toda theory determines the mass of new fields, as we will see in \S\,\ref{sec:check}.
Also, the condition (\ref{zeros-1}) means that some factors in the denominator of Toda correlation function are factored out, and then the rank of gauge group of corresponding 4-dim theory becomes smaller. 
Therefore, it is natural to consider that the setting of new poles (\ref{zeros-1}) corresponds to the setting of D6-branes.
In fact, as we will see in \S\,\ref{sec:check}, the order of these poles in the whole of Toda correlation function is always equal to the number of times which D6-branes pass through NS5-branes, or equivalently, Hanany-Witten transition occurs.

Next, we pay attention to the factors 
in the denominator other than the zero factors (\ref{zeros-1}),
{\em i.e.}~$\Upsilon\left(\upsilon_{j_1,j_2}\right)$ with $(j_1,j_2)\neq (\hj_1,\hj_2)$.
As in the previous case ({\em i.e.}~$s_k$, $\tilde s_k$ are integers),
the cancellation of these factors  and some factors in the numerator may occur. However, it should be in some different manner, so let us now list the ways of cancellation. 
\begin{description}
\item[Case 1] ~For $j_1>\hj_1$ and $j_2=\hj_2$,
the factors $\Upsilon\left(\upsilon_{j_1,j_2}\right)$ are canceled by corresponding factors in the numerator, since
\ba\label{case1}
\upsilon_{j_1,\hj_2}
\,=\,
(Q\vrho-\valp_1)\cdot\sum_{k=\hj_1}^{j_1-1} \ve_k\,.
\ea

\item[Case 2] ~For $j_1<\hj_1$ and $j_2=\hj_2$,
if $\valp_1\cdot \ve_{\hj_1}=0$,
the factors $\Upsilon\left(\upsilon_{j_1,j_2}\right)$ are canceled by corresponding factors in the numerator, since
\ba\label{case2}
\upsilon_{j_1,\hj_2}
\,=\,
Q-(Q\vrho-\valp_1)\cdot\sum_{k=j_1}^{\hj_1} \ve_k\,.
\ea
Here we note that $\Upsilon(x)=\Upsilon(Q-x)$ for ${}^\forall x\in\bbC$.

\item[Case 3] ~For $j_1<\hj_1$ and $j_2=\hj_2$,
if $\valp_1$ is a momentum of a propagator ({\em i.e.}~internal line),
the factors $\Upsilon\left(\upsilon_{j_1,j_2}\right)$ are canceled by corresponding factors in the numerator, since
\ba\label{case4}
\upsilon_{j_1,\hj_2}
\,=\,
\bigl(Q\vrho-(2Q\vrho-\valp_1)\bigr)\cdot\sum_{k=j_1}^{\hj_1-1} \ve_k\,.
\ea
Here we note that a propagator momentum $\valp$ appears in two 3-point functions, like $C(\,*\,,\valp;\,*\,)$ and $C(2Q\vrho-\valp,\,*\,;\,*\,)$,
as we will see in eq.\,(\ref{tree}).

\item[Case 4] ~For $j_1<\hj_1$ and $j_2=\hj_2+1$,
if $\valp_2\cdot \ve_{\hj_2}=0$,
the factors $\Upsilon\left(\upsilon_{j_1,j_2}\right)$ are canceled by corresponding factors in the numerator, since
\ba\label{case5}
\upsilon_{j_1,\hj_2+1}
\,=\,
Q-(Q\vrho-\valp_1)\cdot\sum_{k=j_1}^{\hj_1-1} \ve_k\,.
\ea
\end{description}

There are other ways of cancellation, but we will consider only these four cases in the following discussion.
We note that all these discussions are still valid, if we interchange the indices $1\leftrightarrow 2$, of course.
By putting all results together, we finally find that the 3-point function (\ref{3pt-s}) appearing in AGT-W relation becomes
\ba\label{3pt-c}
C(\valp_1,\valp_2;\varkap \vome_{N-1})
\!&=&\!
\left[\pi\mu\gamma(b^2)b^{2-2b^2}\right]^{(2Q\vrho-\valp_1-\valp_2-\varkappa\vome_{N-1})\cdot\vrho/b}
\\&&\times
\frac{\,\Upsilon(b)^{N-1}\Upsilon(\varkap)\prod'_{e_{1,2}>0}\Upsilon\left((Q\vrho-\valp_1)\cdot\ve_1\right)\Upsilon\left((Q\vrho-\valp_2)\cdot\ve_2\right)\,}{\prod_{j_{1,2}\neq \hj_{1,2}}
\Upsilon\left(\varkap/N+(\valp_1-Q\vrho)\cdot\vlam_{j_1}+(\valp_2-Q\vrho)\cdot\vlam_{j_2}\right)}
\nn
\ea
after the cancellation discussed above.
Here, $\prod_{j_{1,2}\neq \hj_{1,2}}$ in the denominator means 
the product of only the elements with 
\ba
\{j_1,j_2\,|\,(j_1\neq \hj_1,\,\hj_1+1)\cap(j_2\neq \hj_2,\,\hj_2+1)\}
\ea
for all the sets $\{\hj_1,\hj_2\}$ satisfying the condition (\ref{zeros-1}).
And, $\prod'_{e_{1,2}>0}$ in the numerator means 
the product of only the elements with 
\ba\label{cond-e1}
\ve_1\neq 
\begin{cases}
\sum_{k=j_1}^{\hj_1}\ve_k
~~\text{or}~ \sum_{k=j_1}^{\hj_1-1}\ve_k
~~&\,\text{for}~~j_1=1,\cdotsb,\hj_1-1\\
\sum_{k=\hj_1}^{j_1-1}\ve_k &\,\text{for}~~ j_1=\hj_1+1,\cdotsb,N\,.
\end{cases}
\ea
For the former case, it depends on the way of cancellation (case 2\,--\,4).
For $\ve_2$, we should replace all the indices 1 with 2 in eq.\,(\ref{cond-e1}).

\section{Check of AGT-W relation}
\label{sec:check}

Based on the discussions in the previous section, now we begin the check of AGT-W relation by direct calculations.

\subsection{Ansatz for Toda momentum}
\label{sec:ansatz}

Let us now see the diagram (\ref{diagram}) of the whole correlation function again:
\ba\label{diagram1}
{\setlength{\unitlength}{1cm}
\begin{picture}(5,1.7)
 \put(0,0){\line(1,0){2.5}}
 \put(3.5,0){\line(1,0){2.5}}
 \put(1,0){\line(0,1){1}}
 \put(2,0){\line(0,1){1}}
 \put(4,0){\line(0,1){1}}
 \put(5,0){\line(0,1){1}}
 \put(-0.7,-0.1){\small$\vbet_\infty$}
 \put(0.8,1.2){\small$\vbet_{n+1}$}
 \put(1.8,1.2){\small$\vbet_1$}
 \put(3.8,1.2){\small$\vbet_{n-1}$}
 \put(4.8,1.2){\small$\vbet_n$}
 \put(6.2,-0.1){\small$\vbet_0$}
 \put(1.3,0.15){\small$\valp_1$}
 \put(4.3,0.15){\small$\valp_n$}
 \put(2.77,0.2){$\cdots$}
\end{picture}}
\ea
In this paper, we consider only the correlation function with descendant level 0\:\!:
\ba\label{tree}
&&\back\!\!\!
V_{\emp}(\vbet_\infty,\vbet_{n+1},\vbet_1,\cdotsb,\vbet_n,\vbet_0;\valp_1,\cdotsb,\valp_n)
\\[+2pt]&&\!\!\!
=C(\vbet_\infty,\valp_1;\vbet_{n+1})\,
C(2Q\vrho-\valp_1,\valp_2;\vbet_1)
\cdots\:\!
C(2Q\vrho-\valp_{n-1},\valp_n;\vbet_{n-1})\,
C(2Q\vrho-\valp_n,\vbet_0;\vbet_n)
\nn
\ea
where the momenta of vertex operators of 2-dim Toda field theory correspond to the quiver gauge group (\ref{quiver}) with the condition (\ref{group-cond}) of 4-dim $\cN=2$ quiver gauge theory as follows:
\begin{itemize}
\item $\vbet_\infty=Q\vrho+i\vbet_\infty'$ : 
This corresponds to a `full' puncture.
\item $\vbet_k=\left(\dfrac{Q}{2}+im_k\right)N\vome_{N-1}$ or $\left(\dfrac{Q}{2}+im_k\right)N\vome_1$
~(for $k=1,\cdotsb,n+1$) :\\[+3pt]
They correspond to `simple' punctures~\cite{Wyllard:2009}.
We choose the former in this paper.
\item $\vbet_0$ : 
This corresponds to a puncture classified by a Young tableau corresponding to the whole of quiver gauge group $SU(d_1)\times \cdots \times SU(d_n)$~\cite{Kanno:2009}.
\item $\valp_j$ 
(for $j=1,\cdotsb,n$) :
They correspond to propagators, but in the weak coupling limit of $SU(d_j)$, the diagram (\ref{diagram1}) is decomposed into two diagrams as below~\cite{Gaiotto:2009}.
This means that $\valp_j$ should correspond to a puncture with a Young tableau for a quiver gauge group $SU(d_1)\times \cdots \times SU(d_{j-1})$~\cite{Drukker:2010}.
\end{itemize}
\ba\label{diagram2}
{\setlength{\unitlength}{1cm}
\begin{picture}(11.5,1.7)
 \put(0,0){\line(1,0){2.5}}
 \put(3.5,0){\line(1,0){1}}
 \put(7.5,0){\line(1,0){1}}
 \put(9.5,0){\line(1,0){2.5}}
 \put(1,0){\line(0,1){1}}
 \put(2,0){\line(0,1){1}}
 \put(4,0){\line(0,1){1}}
 \put(8,0){\line(0,1){1}}
 \put(10,0){\line(0,1){1}}
 \put(11,0){\line(0,1){1}}
 \put(-0.7,-0.1){\small$\vbet_\infty$}
 \put(0.8,1.2){\small$\vbet_{n+1}$}
 \put(1.8,1.2){\small$\vbet_1$}
 \put(3.8,1.2){\small$\vbet_{j-1}$}
 \put(7.8,1.2){\small$\vbet_{j}$}
 \put(9.8,1.2){\small$\vbet_{n-1}$}
 \put(10.8,1.2){\small$\vbet_n$}
 \put(4.7,-0.1){\small$\valp_j$}
 \put(12.2,-0.1){\small$\vbet_0$}
 \put(5.9,-0.1){\small$2Q\vrho-\valp_j$}
 \put(1.3,0.15){\small$\valp_1$}
 \put(10.3,0.15){\small$\valp_n$}
 \put(2.77,0.2){$\cdots$}
 \put(8.77,0.2){$\cdots$}
 \put(5.25,0.5){$\bigotimes$}
\end{picture}}
\ea
\vspace{1mm}\noindent
Here the parameters $\vbet_\infty'$ and $m_k$ 
are real, and the following conditions are satisfied:
\ba\label{momentum-cond}
\sum_{p=1}^N \beta'_{\infty,p}=\sum_{p=1}^N \beta_{0,p}=\sum_{p=1}^N \alpha_{j,p}=0\,.
\ea
There are no more conditions for $\vbet_\infty$ and $\vbet_k$, 
while $\vbet_0$ and $\valp_j$ must satisfy some additional conditions.
Then let us now explain how to determine their concrete forms.

\sssn{Our ansatz for $\vbet_0$}

According to Gaiotto's discussion~\cite{Gaiotto:2009},
a puncture corresponding $\vbet_0$ can be classified by a Young tableau $[l_1,\cdotsb,l_s]$
({\em i.e.}~the number of boxes in $i$-th column is $l_i$).
In fact, for the gauge theory with quiver gauge group $SU(d_1)\times\cdots\times SU(d_n)$ under the condition (\ref{quiver-cond}), 
$\vbet_0$ corresponds to the puncture with Young tableau whose number of boxes in $j$-th line is $d_{j}-d_{j+1}$. That is,
\ba
[l_1,\cdotsb,l_s]=[d_1-d_2,\cdotsb,d_{n-1}-d_n,d_n]^T
\ea
where ${}^T$ means the transposition of a Young tableau,
and then $s=d_n$.
Note that the total number of boxes is always $N\,(=d_1)$.
Especially, a puncture with tableau $[1^N]$ is called a `full' puncture, and a puncture with tableau $[N-1,1]$ is called a `simple' puncture.

Our ansatz in \cite{Kanno:2009} gives how to determine the form of $\vbet_0$
by using this Young tableau $Y=[l_1,\cdotsb,l_s]$ as follows:
First, we divide $\vbet_0$ into its real and imaginary parts as 
\ba\label{ansatz-beta0}
\vbet_0=Q\vrho_Y+i\vbet_0'\,.
\ea
For the real part, $\vrho_Y$ is defined as
\ba\label{rhoY}
\vrho_Y\!&:=&\!
\vrho-(\vrho_{l_1}\oplus\cdots\oplus\vrho_{l_i}\oplus\cdots\oplus\vrho_{l_s})
\\&=&\!
\bigl(\underbrace{\tfrac{N-l_1}{2},\cdotsb,\tfrac{N-l_1}{2}}_{l_1},\cdotsb,
\underbrace{\tfrac{N-l_i-2\sum_{j=1}^{i-1} l_j}{2},\cdotsb,\tfrac{N-l_i-2\sum_{j=1}^{i-1}l_j}{2}}_{l_i},
\cdotsb,
\underbrace{\tfrac{-N+l_s}{2},\cdotsb,\tfrac{-N+l_s}{2}}_{l_s}\bigr)
\nn
\ea
where
\ba
\vrho_k=\frac12(k-1,k-3,\cdotsb,3-k,1-k)\,.
\ea
For the imaginary part, $\vbet_0'$ is defined as
\ba\label{vbet0}
\vbet_0'=(\underbrace{\beta_{0,1}',\cdotsb,\beta_{0,1}'}_{l_1},\cdotsb,
\underbrace{\beta_{0,i}',\cdotsb,\beta_{0,i}'}_{l_i},\cdotsb,
\underbrace{\beta_{0,s}',\cdotsb,\beta_{0,s}'}_{l_s})
=:\vbet'_{0\:\![l_1,\cdotsb,l_i,\cdotsb,l_s]}\,.
\ea
Therefore, we can find that $\vbet_0\cdot\ve_k=0$ is satisfied for ${}^\forall k\neq l_1+l_2+\cdotsb+l_i$ for ${}^\forall i=1,\cdotsb, s$. 
Note that we don't fix the order of $l_1,\cdotsb,l_s$ at this moment,
but in the following discussion,
we will consistently fix it as $l_1\leq \cdots\leq l_s$.


\sssn{Ansatz for $\valp_j$ by Drukker and Passerini}

According to Gaiotto's discussion~\cite{Gaiotto:2009} again,
in the weak coupling limit of $SU(d_j)$,
{\em i.e.}~when the diagram is decomposed like eq.\,(\ref{diagram2}),
$\valp_j$ should correspond to a puncture for a quiver gauge group $SU(d_1)\times \cdots \times SU(d_{j-1})$.
Therefore, we can use our ansatz (\ref{ansatz-beta0}) here again.
Then the ansatz for $\valp_j$ can be written as
\ba\label{alpj}
\valp_j=Q\vrho_Y+i\left[(\valp'_j,\vec 0)+\vgam_{j\:\![d_j,l_{d_j+1},\cdotsb,l_s]}\right]
\ea
where $Y=[l_1,\cdotsb,l_s]$ is a Young tableau for a quiver gauge group $SU(d_1)\times\cdots\times SU(d_{j-1})$,
so $s=d_{j-1}$ here.
Since the condition (\ref{quiver-cond}) means $d_{j-1}-(d_{j-2}-d_{j-1})\geq d_j$,
if we put in order as $l_1\leq \cdots\leq l_s$, 
we find that $l_1=\cdots=l_{d_j}=1$ is always satisfied.
Then we can put a traceless $d_j$-component vector $\valp_j'$ as in eq.\,(\ref{alpj}), which can be regarded as an $SU(d_j)$ propagator.
The next $\vec 0$ is a $(N-d_j)$-component zero vector.
The remaining part $\vgam_j$ should be determined so that the whole imaginary part $\bigl[(\valp'_j,\vec 0)+\vgam_j\bigr]$ is of the same form as $\vbet'_{0\:\![l_1,\cdotsb,l_s]}$ in eq.\,(\ref{vbet0}). 
In fact, eq.\,(\ref{alpj}) satisfies it, since
\ba\label{valp-im}
&&\back
(\valp'_j,\vec 0)+\vgam_{j\:\![d_j,l_{d_j+1},\cdotsb,l_s]}
\\&&=
(\underbrace{\alpha'_{j,1}}_{l_1=1},\cdotsb,
\underbrace{\alpha'_{j,d_j}}_{l_{d_j}=1},
\underbrace{0,\cdotsb,0}_{N-d_j})
+(\underbrace{\gamma_{j,d_j},\cdotsb,\gamma_{j,d_j}}_{l_1+\cdots+l_{d_j}=d_j},
\underbrace{\gamma_{j,d_j+1},\cdotsb,\gamma_{j,d_j+1}}_{l_{d_j+1}},\cdotsb,
\underbrace{\gamma_{j,s},\cdotsb,\gamma_{j,s}}_{l_s})\,.\nn
\ea
This exactly agrees with the ansatz by Drukker and Passerini~\cite{Drukker:2010}.
Note that $\valp_j\cdot\ve_k=0$ is satisfied for ${}^\forall k\neq l_1+l_2+\cdotsb+l_i$ for ${}^\forall i=1,\cdotsb, s$, just as for $\vbet_0$.

\paragraph{}
Up to now, the proof of AGT-W relation by direct calculations has been done in \cite{Wyllard:2009,Drukker:2010,Kanno:2010}, but it seems to be still restrictive.
Our calculation in the following is also restrictive in that we consider only the correlation functions of $A_{N-1}$ Toda theory with descendant level 0 and the corresponding 1-loop part of partition functions of $SU(N)$ quiver gauge theory. However, we consider a general case of $SU(N)$ quiver gauge group, which is a new point of this paper.
In the remainder of this section, we first consider the simple cases of $A_2$ and $A_3$ Toda theory in \S\,\ref{sec:A2} and \S\,\ref{sec:A3}, 
and then we discuss a general case of $A_{N-1}$ Toda theory in \S\,\ref{sec:AN-1}.

\subsection{Case of $A_2$ Toda theory}
\label{sec:A2}

We have already discussed this case in our previous paper~\cite{Kanno:2010}, but here let us make some modifications.
Especially, we consider all factors in the correlation function (\ref{tree}),
although the unrelated factors to 1-loop partition function 
have been usually neglected in the previous researches.

\sss{For $SU(3)^n$ quiver}
\label{sec:3}

In this case, $\vbet_0$ corresponds to a `full' puncture, then we set
\ba\label{set-3}
\valp_j=Q\vrho+i\valp_j'\,,\quad
\vbet_0=Q\vrho+i\vbet_0'\,,\quad
\vbet_k=\left(\frac{Q}{2}+im_k\right)3\vome_2\,,
\ea
where $j=1,\cdotsb,n$ and $k=1,\cdotsb,n+1$.
Here no conditions other than eq.\,(\ref{momentum-cond}) are imposed.
As we discussed in \S\,\ref{sec:3pt-func}, we note that there never be any zeros in the denominator of all the 3-point functions, since
\ba
\upsilon_{\zeta_1,\zeta_2}=\frac{Q}{2}+i\left(\text{a linear combination of } \alpha'_{j,p},\,\beta'_{0,q},\,m_k\right)
\qquad\!
\text{for}~{}^\forall \zeta_1,\zeta_2=1,2,3
\ea
where $p,q=1,2,3$, and $\upsilon_{\zeta_1,\zeta_2}$ has been defined in eq.\,(\ref{zeros-1}).
Then we require that there should be no zeros also in the numerator.
This requirement (\ref{cond-notail}) means that 
\ba
\alpha'_{j,p}\neq \alpha'_{j,q}\,,\quad
\beta'_{0,p}\neq \beta'_{0,q}\qquad
\text{for}~~p\neq q\,.
\ea
Let us now compare the level-0 correlation function (\ref{tree}), denoted by $V_\emp$, with the 1-loop partition function of gauge theory (\ref{Z1lp}), denoted by $Z_{\text{1-loop}}$.
Then we find that when the correspondence of parameters is set as
\ba\label{mass-3}
\text{$SU(3)$ adjoint scalar VEV}&&
{\vec{\hat a}}_j=i\valp_j' \quad\text{(for $j=1,\cdotsb,n$)}
\nn\\[+2pt]
\text{$SU(3)$ bifundamental mass}&&
\nu_k=\frac{Q}{2}+im_k \quad\text{(for $k=1,\cdotsb,n-1$)}
\nt
\text{$SU(3)$ fundamental mass}&&
\mu_p=\frac{Q}{2}+im_n\pm i\beta'_{0,p}
\quad\text{(for $p=1,2,3$)}
\nt
\text{$SU(3)$ antifundamental mass}&& 
\bar\mu_p=\frac{Q}{2}-im_{n+1}\mp i\beta'_{\infty,p}
\nn\\[+2pt]
\text{Nekrasov's parameters}&& 
\eps_1=b\,,\quad \eps_2=b^{-1}\,,
\ea
we can show that the correlation function can be written as
\ba\label{AGT}
V_\emp\,=\,
A^{n+1}\,h(2Q\vrho)^n\,g(\vbet_\infty)\,g(\vbet_0)
\prod_{k=1}^{n+1} f_2(m_k)
\prod_{j=1}^n 
\prod_{p<q}\,(\alp'_{j,p}-\alp'_{j,q})^2\,
|Z_\text{1-loop}|^2
\ea
where (for $A_{N-1}$ Toda theory)
\ba\label{prod-e}
A\!&:=&\!
\left[\pi\mu\gamma(b^2)b^{2-2b^2}\right]^{2Q|\vrho|^2/b}
\Upsilon(b)^{N-1}
\nt
f_{N-1}(m)\!&:=&\!\left[\pi\mu\gamma(b^2)b^{2-2b^2}\right]^{-(\frac12Q+im)N\vome_{N-1}\cdot\vrho/b}\Upsilon(N(\tfrac12Q+im))
\nt
g(\vbet)\!&:=&\!\left[\pi\mu\gamma(b^2)b^{2-2b^2}\right]^{-\vbet\cdot\vrho/b}
{\prod_{e>0}}'\,\Upsilon((Q\vrho-\vbet)\cdot\ve)
\nt
h(\valp)\!&:=&\!\left[\pi\mu\gamma(b^2)b^{2-2b^2}\right]^{-\valp\cdot\vrho/b}
\ea
where $\prod'_{e>0}$ has been already defined in eq.\,(\ref{3pt-c}).
In the present case, it is equivalent to the usual $\prod_{e>0}$,
since any cancellations discussed in \S\ref{sec:3pt-func} don't occur.
However, if such cancellations occur, we should take the product $\prod'_{e>0}$ following the condition (\ref{cond-e1}).
We will explain it clearly in the following cases.

We finally note that 
$\vome_2$ in the setting of $\vbet_k$ (\ref{set-3}) can be replaced by $\vome_1$.
This replacement slightly changes the parameter correspondence (\ref{mass-3})\:\!: the upper signs are for $\vome_2$, while the lower signs are for $\vome_1$.
The coefficient functions in eq.\,(\ref{AGT}) is also changed:
for the choice of $\vome_1$ instead of $\vome_{N-1}$ in $A_{N-1}$ Toda theory, the factor $f_{N-1}$ must be replaced as 
\ba
f_{N-1}(m)\to
f_1(m):=\left[\pi\mu\gamma(b^2)b^{2-2b^2}\right]^{-(\frac12Q+im)N\vome_1\cdot\vrho/b}\Upsilon(N(\tfrac12Q+im))\,.
\ea

Therefore, we can successfully show that the level-0 correlation function of Toda theory properly corresponds to the 1-loop partition function of $SU(3)^n$ quiver gauge theory.

\sss{For $SU(3)^{n-1}\times SU(2)$ quiver}
\label{sec:32}

In this case, $\vbet_0$ becomes a `simple' puncture.
In the weak coupling limit of the last $SU(2)$, $\valp_n$ becomes a `full' puncture. Then we should change a part of the setting (\ref{set-3}) as
\ba\label{set-32}
\valp_n=Q\vrho+i\left[(\valp_n',0)-3\tilde m_n\vome_2\right]\,,\quad
\vbet_0=\left(\frac{Q}{2}+im_0\right)3\vome_1\,,
\ea
where $\valp'_n$ is a traceless 2-component vector.
The last term in $\valp_n$ did not appear in our previous paper~\cite{Kanno:2010}, but we add here by following the ansatz (\ref{alpj}).

As we discussed in \S\,\ref{sec:3pt-func}, 
there can be the zeros in the denominator of the last 3-point function $C(2Q\vrho-\valp_n,\vbet_0;\vbet_n)$, since
\ba
\ups_{3,2}=i(m_n-m_0-2\tilde m_n)\,.
\ea
This means that the factor $\Ups(\ups_{3,2})$ makes a new pole,
when we set $m_n=m_0+2\tilde m_n$.
Then some of the other factors in the denominator cancel out some factors in the numerator in the way of case 1\,--\,4, or eq.\,(\ref{case1})\,--\,(\ref{case5}).
These cancellations can be summarized as table~\;\!\!\ref{tab:32}.
\begin{table}[t]
\begin{center}
\begin{tabular}{|c||c|rl|}
\hline
case & denominator & \multicolumn{2}{|c|}{numerator} \\\hline\hline
$-$ & $\ups_{3,2}$ & \multicolumn{2}{|c|}{$-\!\!-$}\\\hline
1 & $\ups_{3,3}$ &
 $(Q\vrho-\vbet_0)\,\cdot$&\back\,$\ve_2$\\\hline
2 & $\ups_{3,1}$ &
 $(Q\vrho-\vbet_0)\,\cdot$&\back\,$(\ve_1+\ve_2)$\\\hline
3 & $\ups_{1,2}$ &
 $(Q\vrho-\valp_n)\,\cdot$&\back\,$(\ve_1+\ve_2)$\\\cline{2-4}
&$\ups_{2,2}$ &
 $(Q\vrho-\valp_n)\,\cdot$&\back\,$\ve_2$\\\hline
4 & $\ups_{1,3}$ &
 $(Q\vrho-(2Q\vrho-\valp_n))\,\cdot$&\back\,$(\ve_1+\ve_2)$\\\cline{2-4}
&$\ups_{2,3}$ &
 $(Q\vrho-(2Q\vrho-\valp_n))\,\cdot$&\back\,$\ve_2$\\\hline
\end{tabular}
\caption{Cancellations in the case of $SU(3)^{n-1}\times SU(2)$ quiver}
\label{tab:32}
\end{center}
\end{table}
After the cancellations, the remaining factors in the denominator are 
\ba
\Ups(\ups_{\zeta,1})=\Ups\left(\frac{Q}{2}+i(-\alpha'_{n,\zeta}+3m_0+3\tilde m_n)\right)
\quad~\text{for}~~\zeta=1,2
\ea
from which we can read off the mass of a $SU(2)$ fundamental matter field, by comparing with the 1-loop partition function of gauge theory.
Therefore, we find the correspondence of parameters as follows:
\ba\label{mass-32}
\text{$SU(2)$ adjoint scalar VEV}&&
{\vec{\hat a}}_n=i\valp'_n
\nn\\[+2pt]
\text{$SU(3)\times SU(2)$ bifundamental mass}&&
\nu_{n-1}=\frac{Q}{2}+im_{n-1}
\nt
\text{$SU(3)$ fundamental mass}&&
\mu^{(3)}=\frac{Q}{2}+i(m_{n-1}+2\tilde m_n)
\nt
\text{$SU(2)$ fundamental mass}&&
\mu^{(2)}=\frac{Q}{2}+i(\pm 3m_0+3\tilde m_n)
\ea
with the condition $m_n=\pm m_0+2\tilde m_n$.
We show here only a different part from the previous case (\ref{mass-3}).
The double signs are related to the choice of $\vome_1$\,/\,$\vome_2$ in the setting of $\vbet_0$ (\ref{set-32}): the upper signs are for $\vome_1$, while the lower signs are for $\vome_2$.
Note that the newly added parameter $\tilde m_n$ makes apart $SU(3)$ fundamental mass from $SU(3)\times SU(2)$ bifundamental mass.

Finally we can show that under the correspondence of parameters (\ref{mass-32}), 
the correlation function can be written as
\ba\label{AGT-32}
V_\emp\,=\,\frac{A^{n+1}}{\Upsilon(0)}\,
h(2Q\vrho)^n\,g(\vbet_\infty)\,g(\vbet_0)
\prod_{k=1}^{n+1} f_2(m_k)
\prod_{j=1}^n 
\prod_{p<q}\,(\alp'_{j,p}-\alp'_{j,q})^2\,
|Z_\text{1-loop}|^2
\ea
where $p,q=1,2,3$ for $j=1,\cdotsb,n-1$ and $p,q=1,2$ for $j=n$.
Note that when we take the product $\prod'_{e>0}$ in $g(\vbet_0)$ defined in eq.\,(\ref{prod-e}),
the factors of numerator listed in table~\!\!\;\ref{tab:32},
{\em i.e.}~$(Q\vrho-\vbet_0)\cdot \ve_2$ and  $(Q\vrho-\vbet_0)\cdot(\ve_1+\ve_2)$, must be removed.
We also note that the order of pole $1/\Ups(0)$ ({\em i.e.}~one, in this case) is equal to the number of times which Hanany-Witten transition occurs in the D4/NS5/D6-brane system, as we discussed in \S\,\ref{sec:3pt-func}.

\subsection{Case of $A_3$ Toda theory}
\label{sec:A3}

Now we incline to rush into the proof of AGT-W relation in a general case of  $A_{N-1}$ Toda theory,
but in this subsection, we check the correspondence in the case of $A_3$ Toda theory, and pile up more observations and discussions on concrete simple examples.

\sss{For $SU(4)^n$ quiver}
\label{sec:4}

In this case, we can show the correspondence very similarly to
the case of $SU(3)^n$ quiver. 
Since $\vbet_0$ corresponds to a `full' ($[1^4]$) puncture,
the setting of Toda momenta is 
\ba\label{set-4}
\valp_j=Q\vrho+i\valp_j'\,,\quad
\vbet_0=Q\vrho+i\vbet_0'\,,\quad
\vbet_k=\left(\frac{Q}{2}+im_k\right)4\vome_3\,,
\ea
where $j=1,\cdotsb,n$ and $k=1,\cdotsb,n+1$. 
Then the correspondence of parameters becomes
\ba\label{mass-4}
\text{$SU(4)$ adjoint scalar VEV}&&
{\vec{\hat a}}_j=i\valp_j' \quad\text{(for $j=1,\cdotsb,n$)}
\nn\\[+2pt]
\text{$SU(4)$ bifundamental mass}&&
\nu_k=\frac{Q}{2}+im_k \quad\text{(for $k=1,\cdotsb,n-1$)}
\nt
\text{$SU(4)$ fundamental mass}&&
\mu_p=\frac{Q}{2}+im_n\pm i\beta'_{0,p}
\quad\text{(for $p=1,\cdotsb,4$)}
\nt
\text{$SU(4)$ antifundamental mass}&& 
\bar\mu_p=\frac{Q}{2}-im_{n+1}\mp i\beta'_{\infty,p}
\ea
where the lower signs are for the choice of $\vome_1$, instead of $\vome_3$ in the setting of $\vbet_k$.
The final result is of the same form as eq.\,(\ref{AGT}),
so we don't write it down here.

\makeatletter
\newsavebox{\@parc@ption}
\def\parcaption#1{%
\sbox{\@parc@ption}{\shortstack[l]{#1}}%
\setbox\@tempboxa\hbox{\csname fnum@\@captype\endcsname}%
\@tempdima\columnwidth \advance\@tempdima-\wd\@tempboxa
\@tempdimb.8\@tempdima 
\ifdim\wd\@parc@ption>\@tempdimb \@tempdima\@tempdimb
\else\@tempdima\wd\@parc@ption\fi
\sbox{\@tempboxa}{\parbox[t]{\@tempdima}{#1}}%
\caption{\usebox{\@tempboxa}}}
\makeatother

\sss{For $SU(4)^{n-1}\times SU(3)$ quiver}
\label{sec:43}

In this case, $\vbet_0$ corresponds to a $[2,1,1]$ puncture.
In the weak coupling limit of $SU(3)$, $\valp_n$ becomes a `full' puncture.
Then we should change a part of the setting (\ref{set-4}) as
\ba\label{set-43}
\valp_n=Q\vrho+i\left[(\valp'_n,0)-4\tilde m_n\vome_3\right]\,,\quad
\vbet_0=Q\vrho_{[1,1,2]}+i\vbet'_{0\:\![1,1,2]}\,,
\ea
where $\valp'_n$ is a traceless 3-component vector.
$\vrho_{[1,1,2]}$ 
and $\vbet'_{0\:\![1,1,2]}$ are defined in eq.\,(\ref{rhoY}) and (\ref{vbet0}).
Let us here comment on the other choices of Young tableau,
{\em i.e.}~$Y=[2,1,1]$, $[1,2,1]$\;\!:
The former has no problem, if we redo the discussion in \S\,\ref{sec:ansatz}
for a Young tableau $[l_1,\cdotsb,l_s]$ with $l_1\geq\cdots\geq l_s$
and change our ansatz for $\valp_n$.
The latter, on the other hand, causes a problem, since it cannot have sufficient degrees of freedom of fundamental matter fields.

The remaining part of discussion is parallel to the case of
$SU(3)^{n-1}\times SU(2)$ quiver. 
There can be the zeros in the denominator of the last 3-point function
$C(2Q\vrho-\valp_n,\vbet_0;\vbet_n)$, since
\ba
\ups_{4,3}=i(m_n+\beta'_{0,3}-3\tilde m_n)\,.
\ea
Then the factor $\Ups(\ups_{4,3})$ makes a new pole, 
when we set $m_n=-\beta'_{0,3}+3\tilde m_n$.
\begin{table}[t]
\begin{center}
\begin{tabular}{|c||c|rl|}
\hline
case & denominator & \multicolumn{2}{|c|}{numerator}\\
\hline\hline 
$-$ & $\ups_{4,3}$ & \multicolumn{2}{|c|}{$-\!\!-$}\\\hline
1 & $\ups_{4,4}$ &
 $(Q\vrho-\vbet_0)\,\cdot$&\back\,$\ve_3$\\\hline
2 & $\ups_{4,\zeta_2}$ &
 $(Q\vrho-\vbet_0)\,\cdot$&\back\,$(\ve_{\zeta_2}\!+\cdots+\ve_3)$\\\hline
3 & $\ups_{\zeta_1,3}$ &
 $(Q\vrho-\valp_n)\,\cdot$&\back\,$(\ve_{\zeta_1}\!+\cdots+\ve_3)$\\\hline
4 & $\ups_{\zeta_1,4}$ &
 $(Q\vrho-(2Q\vrho-\valp_n))\,\cdot$&\back\,$(\ve_{\zeta_1}\!+\cdots+\ve_3)$\\\hline
\end{tabular}
\parcaption{Cancellations in the case of $SU(4)^{n-1}\times SU(3)$ quiver\\ 
$(\zeta_1=1,2,3,~\zeta_2=1,2)$}
\label{tab:43}
\end{center}
\end{table}
After the cancellations listed in table~\!\!\;\ref{tab:43},
the remaining factors in the denominator are
\ba
\Ups(\ups_{\zeta_1,\zeta_2})=\Ups\left(\frac{Q}{2}+i
(-\alpha'_{n,\zeta_1}+\beta'_{0,\zeta_2}-\beta'_{0,3}+4\tilde m_n)\right)
\quad~\text{for}~~
\zeta_1=1,2,3\,,~~\zeta_2=1,2
\ea
from which we can read off the mass of two $SU(3)$ fundamental matter fields, by comparing with the 1-loop partition function of gauge theory.
Therefore, by using $\beta'_{0,1}+\beta'_{0,2}+2\beta'_{0,3}=0$, 
we find the correspondence of parameters as
\ba\label{mass-43}
\text{$SU(3)$ adjoint scalar VEV}&&
{\vec{\hat a}}_n=i\valp'_n
\nn\\[+2pt]
\text{$SU(4)\times SU(3)$ bifundamental mass}&&
\nu_{n-1}=\frac{Q}{2}+im_{n-1}
\nt
\text{$SU(4)$ fundamental mass}&&
\mu^{(4)}=\frac{Q}{2}+i(m_{n-1}+3\tilde m_n)
\nt
\text{$SU(3)$ fundamental mass}&&
\mu^{(3)}_{1,2}=\begin{cases}
\frac{Q}{2}+i(\frac32\beta'_{0,1}+\frac12\beta'_{0,2}+4\tilde m_n)\\
\frac{Q}{2}+i(\frac12\beta'_{0,1}+\frac32\beta'_{0,2}+4\tilde m_n)
\end{cases}
\ea
with the condition $m_n=\frac12(\beta'_{0,1}+\beta'_{0,2})+\tilde m_n$.
The final result is of the same form as eq.\,(\ref{AGT-32})
with $p,q=1,\cdotsb,4$ for $j=1,\cdotsb,n-1$ and $p,q=1,2,3$ for $j=n$.

\sss{For $SU(4)^{n-2}\times SU(3)\times SU(2)$ quiver}
\label{sec:432}

In this case, $\vbet_0$ corresponds to a simple ($[3,1]$) puncture.
In the weak coupling limit of $SU(2)$,
$\valp_n$ becomes a $[2,1,1]$ puncture.
In the weak coupling limit of $SU(3)$,
$\valp_{n-1}$ becomes a `full' puncture.
Then we should change a part of the setting (\ref{set-4}) as
\ba\label{set-432}
&&\back
\valp_n=Q\vrho_{[1,1,2]}+i\left[(\valp_n',0,0)-4\tilde m_n\vome_2\right]\,,\quad
\vbet_0=\left(\frac{Q}{2}+im_0\right)4\vome_1\,,
\nt&&\back
\valp_{n-1}=Q\vrho+i\left[(\valp_{n-1}',0)-4\tilde m_{n-1}\vome_3\right]\,,
\ea
where 
$\valp_n'$ is a traceless 2-component vector,
$\valp_{n-1}'$ is a traceless 3-component vector, and
$\vome_2=(1,1,-1,-1)/2$.

Now we study the zeros in the denominator of 3-point functions.
On the second last function $C(2Q\vrho-\valp_{n-1},\valp_n;\vbet_{n-1})$, we can reuse the results of the previous case of $SU(4)^{n-1}\times SU(3)$ quiver only after a reparametrization,
{\em e.g.}~$(m_n, \tilde m_n, \beta'_{0,3}) \to (m_{n-1}, \tilde m_{n-1}, 2\tilde m_n)$.
Then we find the condition 
$m_{n-1}=-2\tilde m_n+3\tilde m_{n-1}$. 
Next we consider the last 3-point function $C(2Q\vrho-\valp_n,\vbet_0;\vbet_n)$.
There can be the zeros in the denominator, since
\ba
\ups_{4,3}=\ups_{3,2}=i(m_n-m_0-2\tilde m_n)\,.
\ea
Then the factors $\Ups(\ups_{4,3})$ and $\Ups(\ups_{3,2})$ make new two poles, 
when we set $m_n=m_0+2\tilde m_n$.
\begin{table}[t]
\begin{center}
\begin{tabular}{|c||c|rl|}
\hline
case & denominator & \multicolumn{2}{|c|}{numerator} \\\hline\hline
$-$ & $\ups_{4,3}$ & \multicolumn{2}{|c|}{$-\!\!-$}\\\hline
1 & $\ups_{4,4}$ &
 $(Q\vrho-\vbet_0)\,\cdot$&\back\,$\ve_3$\\\hline
2 & $\ups_{4,\zeta_2}$ &
 $(Q\vrho-\vbet_0)\,\cdot$&\back\,$(\ve_{\zeta_2}\!+\cdots+\ve_3)$\\\hline
4 & $\ups_{\zeta_1,4}$ &
 $(Q\vrho-(2Q\vrho-\valp_n))\,\cdot$&\back\,$(\ve_{\zeta_1}\!+\cdots+\ve_3)$\\\hline\hline
$-$ & $\ups_{3,2}$ & \multicolumn{2}{|c|}{$-\!\!-$}\\\hline
1 & $\ups_{3,3}$ &
 $(Q\vrho-\vbet_0)\,\cdot$&\back\,$\ve_2$\\\hline
2 & $\ups_{3,1}$ &
 $(Q\vrho-\vbet_0)\,\cdot$&\back\,$(\ve_1+\ve_2)$\\\hline
3 & $\ups_{\zeta_1',2}$ &
 $(Q\vrho-\valp_n)\,\cdot$&\back\,$(\ve_{\zeta_1'}\!+\cdots+\ve_2)$\\\hline
4 & $\ups_{\zeta_1',3}$ &
 $(Q\vrho-(2Q\vrho-\valp_n))\,\cdot$&\back\,$(\ve_{\zeta_1'}\!+\cdots+\ve_2)$\\\hline
\end{tabular}
\parcaption{Cancellations in the case of $SU(4)^{n-2}\times SU(3)\times SU(2)$ quiver\\
$(\zeta_1=1,2,3,~\zeta_2=1,2,~\zeta_1'=1,2)$}
\label{tab:432}
\end{center}
\end{table}
After the cancellations listed in table~\:\!\!\ref{tab:432}, the remaining factors in the denominator are
\ba
\Ups(\ups_{\zeta,1})=\Ups\left(\frac{Q}{2}+i(-\alpha'_{n,\zeta}+m_n+3m_0+2\tilde m_n)\right)
\quad~\text{for}~~\zeta=1,2
\ea
from which we can read off the mass of a $SU(2)$ fundamental matter fields,
by comparing with the 1-loop partition function of gauge theory.
Therefore, we find the correspondence of parameters as 
\ba\label{mass-432}
\text{$SU(3)$ adjoint scalar VEV}&&
{\vec{\hat a}}_{n-1}=i\valp'_{n-1}
\nn\\[+4pt]
\text{$SU(2)$ adjoint scalar VEV}&&
{\vec{\hat a}}_n=i\valp'_n
\nn\\[+2pt]
\text{$SU(4)\times SU(3)$ bifundamental mass}&&
\nu_{n-2}=\frac{Q}{2}+im_{n-2}
\nt
\text{$SU(3)\times SU(2)$ bifundamental mass}&&
\nu_{n-1}=\frac{Q}{2}+im_{n-1}
\nt
\text{$SU(4)$ fundamental mass}&&
\mu^{(4)}=\frac{Q}{2}+i(m_{n-2}+3\tilde m_{n-1})
\nt
\text{$SU(2)$ fundamental mass}&&
\mu^{(2)}=\frac{Q}{2}+i(4m_0+4\tilde m_n)
\ea
with the conditions $m_{n-1}=-2\tilde m_n+3\tilde m_{n-1}$ and $m_n=m_0+2\tilde m_n$.

Finally we can show that under the correspondence of parameters (\ref{mass-432}), 
the correlation function can be written as
\ba\label{AGT-432}
V_\emp\,=\,\frac{A^{n+1}}{\Upsilon(0)^3}\,
h(2Q\vrho)^n\,g(\vbet_\infty)\,g(\vbet_0)
\prod_{k=1}^{n+1} f_3(m_k)
\prod_{j=1}^n 
\prod_{p<q}\,(\alp'_{j,p}-\alp'_{j,q})^2\,
|Z_\text{1-loop}|^2
\ea
where $p,q=1,\cdotsb,4$ for $j=1,\cdotsb,n-2$\,;
$p,q=1,2,3$ for $j=n-1$\,; 
$p,q=1,2$ for $j=n$.
Here we repeat the comments for the case of $SU(3)^{n-1}\times SU(2)$ quiver:
The product $\prod'_{e>0}$ in $g(\vbet_0)$ must be taken with the factors of numerator in table~\;\!\!\ref{tab:432} removed.
The order of pole $1/\Ups(0)$ ({\em i.e.}~three, in this case) is equal to the total number of times which Hanany-Witten transition occurs in the D4/NS5/D6-brane system.

\sss{For $SU(4)^{n-1}\times SU(2)$ quiver}
\label{sec:42}

In this case, $\vbet_0$ corresponds to a $[2,2]$ puncture.
In the weak coupling limit of $SU(2)$, $\valp_n$ becomes a `full' puncture. Then we should change a part of the setting (\ref{set-4}) as
\ba\label{set-42}
\valp_n=Q\vrho+i\left[(\valp'_n,0,0)+\vgam_{n\:\![2,1,1]}\right]\,,\quad
\vbet_0=\left(\frac{Q}{2}+im_0\right)4\vome_2\,,
\ea
where $\valp'_n$ is a traceless 2-component vector,
and $\vgam_n$ is of the form $(\vgam_{n,1},\vgam_{n,1},\vgam_{n,2},\vgam_{n,3})$.

\begin{table}[t]
\begin{center}
\begin{tabular}{|c||c|rl|}
\hline
case & denominator & \multicolumn{2}{|c|}{numerator} \\\hline\hline
$-$ & $\ups_{4,3}$ & \multicolumn{2}{|c|}{$-\!\!-$}\\\hline
1 & $\ups_{4,4}$ &
 $(Q\vrho-\vbet_0)\,\cdot$&\back\,$\ve_3$\\\hline
2 & $\ups_{4,\zeta_2}$ &
 $(Q\vrho-\vbet_0)\,\cdot$&\back\,$(\ve_{\zeta_2}\!+\cdots+\ve_3)$\\\hline
3 & $\ups_{\zeta_1,3}$ &
 $(Q\vrho-\valp_n)\,\cdot$&\back\,$(\ve_{\zeta_1}\!+\cdots+\ve_3)$\\\hline
4 & $\ups_{\zeta_1,4}$ &
 $(Q\vrho-(2Q\vrho-\valp_n))\,\cdot$&\back\,$(\ve_{\zeta_1}\!+\cdots+\ve_3)$
\\\hline\hline
$-$ & $\ups_{3,1}$ & \multicolumn{2}{|c|}{$-\!\!-$}\\\hline
1 & $\ups_{3,2}$ &
 $(Q\vrho-\vbet_0)\,\cdot$&\back\,$\ve_2$\\\hline
3 & $\ups_{\zeta_1',1}$ &
 $(Q\vrho-\valp_n)\,\cdot$&\back\,$(\ve_{\zeta_1'}\!+\cdots+\ve_2)$\\\hline
4 & $\ups_{\zeta_1',2}$ &
 $(Q\vrho-(2Q\vrho-\valp_n))\,\cdot$&\back\,$(\ve_{\zeta_1'}\!+\cdots+\ve_2)$\\\hline
\end{tabular}
\parcaption{Cancellations in the case of $SU(4)^{n-1}\times SU(2)$ quiver\\
$(\zeta_1=1,2,3,~\zeta_2=1,2,~\zeta_1'=1,2)$}
\label{tab:42}
\end{center}
\end{table}
As we discussed repeatedly, there can be the zeros in the denominator of the last 3-point function $C(2Q\vrho-\valp_n,\vbet_0;\vbet_n)$, since
\ba
\ups_{4,3}=i(m_n-2m_0-\gamma_{n,3})\,,\quad
\ups_{3,1}=i(m_n+2m_0-\gamma_{n,2})\,.
\ea
Then the factors $\Ups(\ups_{4,3})$ and $\Ups(\ups_{3,1})$ make new poles, when we set $m_n=2m_0+\gamma_{n,3}=-2m_0+\gamma_{n,2}$.
After the cancellations listed in table~\:\!\!\ref{tab:42}, we find that there are no remaining factors in the denominator. This is consistent with that there are no $SU(2)$ fundamental matter fields in the corresponding gauge theory.
Then the correspondence of parameters is
\ba\label{mass-42}
\text{$SU(2)$ adjoint scalar VEV}&&
{\vec{\hat a}}_n=i\valp'_n
\nn\\[+2pt]
\text{$SU(4)\times SU(2)$ bifundamental mass}&&
\nu_{n-1}=\frac{Q}{2}+im_{n-1}
\nt
\text{$SU(4)$ fundamental mass}&&
\mu^{(4)}_{1,2}=
\begin{cases}
\frac{Q}{2}+i(m_{n-1}+m_n+2m_0)\\
\frac{Q}{2}+i(m_{n-1}+m_n-2m_0)
\end{cases}
\ea
Therefore, we can show that under this correspondence of parameters,
the correlation function can be written as
\ba\label{AGT-42}
V_\emp\,=\,\frac{A^{n+1}}{\Upsilon(0)^2}\,
h(2Q\vrho)^n\,g(\vbet_\infty)\,g(\vbet_0)
\prod_{k=1}^{n+1} f_3(m_k)
\prod_{j=1}^n 
\prod_{p<q}\,(\alp'_{j,p}-\alp'_{j,q})^2\,
|Z_\text{1-loop}|^2
\ea
where $p,q=1,\cdotsb,4$ for $j=1,\cdotsb,n-1$
and $p,q=1,2$ for $j=n$.

\subsection{General case of $A_{N-1}$ Toda theory}
\label{sec:AN-1}

We finally start the proof of AGT-W relation in a general case of $A_{N-1}$ Toda field theory. First, we decompose the original diagram (\ref{diagram1}) into the $(n+1)$ 3-point functions:
\ba\label{diagram3}
{\setlength{\unitlength}{1cm}
\begin{picture}(13.3,1.7)
 \put(0,0){\line(1,0){1.5}}
 \put(5.9,0){\line(1,0){1}}
 \put(11.6,0){\line(1,0){1.5}}
 \put(1,0){\line(0,1){1}}
 \put(6.4,0){\line(0,1){1}}
 \put(12.1,0){\line(0,1){1}}
 \put(0.8,1.2){\small$\vbet_{n+1}$}
 \put(6.25,1.2){\small$\vbet_{j}$}
 \put(11.9,1.2){\small$\vbet_n$}
 \put(-0.6,-0.1){\small$\vbet_\infty$}
 \put(1.7,-0.1){\small$\valp_1$}
 \put(4.3,-0.1){\small$2Q\vrho-\valp_j$}
 \put(7.1,-0.1){\small$\valp_{j+1}$}
 \put(10,-0.1){\small$2Q\vrho-\valp_n$}
 \put(13.25,-0.1){\small$\vbet_0$}
 \put(2.94,0.5){$\cdots$}
 \put(8.64,0.5){$\cdots$}
 \put(2.2,0.5){$\bigotimes$}
 \put(3.7,0.5){$\bigotimes$}
 \put(7.9,0.5){$\bigotimes$}
 \put(9.4,0.5){$\bigotimes$}
\end{picture}}
\ea
These 3-point functions can be classified into four types,
so let us discuss the correspondence to the 1-loop partition function of gauge theory for each type.

\sssn{Type 1\;\!: the first 3-point function $C(\vbet_\infty,\valp_1;\vbet_{n+1})$}

For all the quiver gauge group, the corresponding momenta of Toda vertex operators are 
\ba\label{set-N1}
\vbet_\infty=Q\vrho+i\vbet'_\infty\,,\quad
\valp_1=Q\vrho+i\valp'_1\,,\quad
\vbet_{n+1}=\left(\frac{Q}{2}+im_{n+1}\right)N\vome_{N-1}\,.
\ea
Then, just as in the case of $SU(3)^n$ and $SU(4)^n$ quiver, 
there never be any zeros in the denominator, since 
\ba
\ups_{\zeta_1,\zeta_2}=\frac{Q}{2}+i(\beta'_{\infty,\zeta_1}+\alpha'_{1,\zeta_2}+m_{n+1})
\qquad\text{for}~{}^\forall \zeta_1,\zeta_2=1,\cdotsb,N\,.
\ea
Moreover, no cancellations of the factors in the denominator and those in the numerator occur. Therefore, if we set the correspondence of parameters as
\ba\label{mass-N1}
\text{$SU(N)$ adjoint scalar VEV}&&
\vha_1=i\valp_1'
\nn\\[+2pt]
\text{$SU(N)$ antifundamental mass}&& 
\bar\mu_p=\frac{Q}{2}-im_{n+1}- i\beta'_{\infty,p}
\quad\text{(for $p=1,\cdotsb,N$)}~
\ea
we can show the correspondence of the 3-point function and the 1-loop partition function as
\ba\label{corr-N1}
C(\vbet_\infty,\valp_1;\vbet_{n+1})
~=~
A\,
g(\vbet_\infty) f_{N-1}(m_{n+1}) h(\valp_1)
\prod_{\brp=1}^N\,\bigl|z^\tx{1lp}_\tx{afd}(\vha_1,\bmu_{\brp})\bigr|^2
\prod_{e>0}\Ups(-i\valp'_1\cdot \ve)
\ea
where $A$, $g$, $f_{N-1}$ and $h$ are defined in eq.\,(\ref{prod-e}),
and $z^\tx{1lp}_\tx{afd}$ is defined in eq.\,(\ref{1lpfactor}).
Note that the last factor, together with the factor from the next 3-point function $C(2Q\vrho-\valp_1,\valp_2;\vbet_1)$, corresponds to the factor $z^\tx{1lp}_\tx{vec}(\vha_1)$ in the 1-loop partition function as
\ba\label{vec}
\prod_{e>0} \Ups(-i\valp'_1\cdot \ve)\Ups(i\valp'_1\cdot \ve)
\!&=&\!
\prod_{e>0}\left|\Gamma_2 (i\valp'_1\cdot \ve)\Gamma_2(Q+i\valp'_1\cdot \ve)\right|^{-2}
\nt&=&\!
\prod_{e>0}\frac{\left|i\valp'_1\cdot\ve\right|^2}{\left|\Gamma_2 (i\valp'_1\cdot \ve+b)\Gamma_2(i\valp'_1\cdot \ve+b^{-1})\right|^2}
\nt&=&\!
\prod_{p<q}\left|\alpha'_{1,p}-\alpha'_{1,q}\right|^2
\bigl|z^\tx{1lp}_\tx{vec}(\vha_1)\bigr|^2\,.
\ea
Here we use the properties of $\Gamma_2$-function shown in eq.\,(\ref{prop-gamma}).

\sssn{Type 2\;\!: 3-point function $C(2Q\vrho-\valp_j,\valp_{j+1};\vbet_j)$ with the rank $d_j=d_{j+1}=N$}

The momenta of Toda vertex operators are set as 
\ba
\valp_j=Q\vrho+i\valp'_j\,,\quad
\valp_{j+1}=Q\vrho+i\valp'_{j+1}\,,\quad
\vbet_j=\left(\frac{Q}{2}+im_j\right)N\vome_{N-1}\,.
\ea
This is just a reparametrization of eq.\,(\ref{set-N1}),
so the discussion is almost parallel to type 1.
Then if the correspondence of parameters is set as
\ba\label{mass-N2}
\text{$SU(N)$ adjoint scalar VEV}&&
\vha_j=i\valp_j'\,,\quad
\vha_{j+1}=i\valp_{j+1}'\,,
\nn\\[+2pt]
\text{$SU(N)$ bifundamental mass}&& 
\nu_j=\frac{Q}{2}+im_j\,,
\ea
we can show the correspondence of 3-point function and the 1-loop partition function as 
\ba\label{corr-N2}
C(2Q\vrho-\valp_j,\valp_{j+1};\vbet_j)
\!&=&\!
A\,
f_{N-1}(m_j) h(2Q\vrho-\valp_j)h(\valp_{j+1})
\nn\\[+3pt]&&\times\,
\bigl|z^\tx{1lp}_\tx{bfd}(\vha_j,\vha_{j+1},\nu_j)\bigr|^2
\prod_{e>0}\Ups(i\valp'_j\cdot \ve)\Ups(-i\valp'_{j+1}\cdot \ve)\,.
\ea
Just as we discussed in eq.\,(\ref{vec}), 
the last two factors correspond to the factors 
$z^\tx{1lp}_\tx{vec}(\vha_j)$ and $z^\tx{1lp}_\tx{vec}(\vha_{j+1})$,
together with the factors from the next 3-point functions.

\sssn{Type 3\;\!: 3-point function $C(2Q\vrho-\valp_j,\valp_{j+1};\vbet_j)$ with the rank $d_j>d_{j+1}$}

We finally discuss the part of descending tail.
According to the ansatz (\ref{alpj}), we set the momenta of Toda vertex operators as 
\ba\label{set-N3}
&&\back
\valp_j=Q\vrho_{Y_j}+i\,\Bigl[(\valp'_j,\vec 0)+\vgam_{j\:\![d_j,l_{d_j+1},\cdotsb,l_s]}\Bigr],\quad
\vbet_{j}=\left(\frac{Q}{2}+im_{j}\right)N\vome_{N-1}\,,\nt
&&\back
\valp_{j+1}=Q\vrho_{Y_{j+1}}+i\,\Bigl[(\valp'_{j+1},\vec 0)+\vgam_{j+1\:\![d_{j+1},l'_{d_{j+1}+1},\cdotsb,l'_{s'}]}\Bigr]\,,
\ea
where $s=d_{j-1}$ and $s'=d_j$. The Young tableaux are
\ba\label{Yj}
Y_j\!&:=&\!
[d_1-d_2,\cdotsb,d_{j-2}-d_{j-1},d_{j-1}]^T
\nt&=&\!
[1^{2d_{j-1}-d_{j-2}},l_x,\cdotsb,l_i,\cdotsb,l_s]
\\
Y_{j+1}\!&:=&\!
[d_1-d_2,\cdotsb,d_{j-2}-d_{j-1},d_{j-1}-d_j,d_j]^T
\nt&=&\!
[1^{2d_j-d_{j-1}},2^{2d_{j-1}-d_j-d_{j-2}},l_x+1,\cdotsb,l_i+1,\cdotsb,l_s+1]
\ea
where $x:=2d_{j-1}-d_{j-2}+1$.

As we saw in \S\,\ref{sec:A2} and \S\,\ref{sec:A3},
the real part of $\ups_{\zeta_1,\zeta_2}$ for ${}^\exists \zeta_1,\,\zeta_2$ becomes zero,
and the imaginary part gives the conditions for momenta.
Let us here check this for a general case.
From the definition (\ref{zeros-1}), $\ups_{\zeta_1,\zeta_2}$ can be written as
\ba\label{ups-N}
\ups_{\zeta_1,\zeta_2}=
Q\left[(\vrho-\vrho_{Y_j})_{\zeta_1}-(\vrho-\vrho_{Y_{j+1}})_{\zeta_2}+\frac12\right]
+i\Bigl[-({\rm Im\,}\valp_j)_{\zeta_1}+({\rm Im\,}\valp_{j+1})_{\zeta_2}+m_j\Bigr]
\ea
where $(\vrho_*)_{\zeta}$ denotes the $\zeta$-th component of a vector $\vrho_*$.
From the definition (\ref{rhoY}), 
\ba
\vrho-\vrho_{Y_j}\!&=&\!\bigl(
\underbrace{0,\cdots\cdotsb,0}_{2d_{j-1}-d_{j-2}},
\underbrace{\tfrac{l_x-1}{2},\cdotsb,\tfrac{1-l_x}{2}}_{l_x},\cdotsb,
\underbrace{\tfrac{l_i-1}{2},\cdotsb,\tfrac{1-l_i}{2}}_{l_i},\cdotsb,
\underbrace{\tfrac{l_s-1}{2},\cdotsb,\tfrac{1-l_s}{2}}_{l_s}
\bigr)
\\[+2pt]
\vrho-\vrho_{Y_{j+1}}\!&=&\!\bigl(
\underbrace{0,\cdotsb,0}_{2d_j-d_{j-1}},
\underbrace{\tfrac12,-\tfrac12,\cdotsb,\tfrac12,-\tfrac12}_{2(2d_{j-1}-d_j-d_{j-2})},
\underbrace{\tfrac{l_x}{2},\cdotsb,\tfrac{-l_x}{2}}_{l_x+1},\cdotsb,
\underbrace{\tfrac{l_i}{2},\cdotsb,\tfrac{-l_i}{2}}_{l_i+1},\cdotsb,
\underbrace{\tfrac{l_s}{2},\cdotsb,\tfrac{-l_s}{2}}_{l_s+1}
\bigr)\,.
\nn
\ea
Therefore, we can find the condition that the real part of $v_{\zeta_1,\zeta_2}$ becomes zero:
\ba\label{cond-zeta}
(\zeta_1,\zeta_2)
=\begin{cases}
(\zeta,\,2\zeta-d_{j-1}-1) &~\text{for}~~\zeta=d_j+1,\cdotsb,2d_{j-1}-d_{j-2}\\
(\zeta,\,\zeta-s-1+i) &~\text{for}~~\zeta-\bigl[(2d_{j-1}-d_{j-2})+\sum_{\ell=x}^{i-1} l_\ell\bigr]=1,\cdotsb,l_i
\end{cases}
\ea
where $i=x,\cdotsb,s$.
Note that this condition can be satisfied for 
${}^\forall \zeta_1=d_j+1,\cdotsb,N$.
Then we have the poles $1/\Ups(0)^{N-d_j}$ in the 3-point function, 
if we impose the following $(N-d_j)$ conditions on the momenta:
\ba\label{cond-N}
m_j=({\rm Im\,}\valp_j)_{\zeta_1}-({\rm Im\,}\valp_{j+1})_{\zeta_2}
\qquad\text{for}~~
(\zeta_1,\zeta_2)\in \text{eq.\,(\ref{cond-zeta})}\,.
\ea
Note that these conditions are not always independent,
as we saw in the case of $SU(4)^{n-2}\times SU(3)\times SU(2)$ quiver. 
This is because 
$({\rm Im\,}\valp_j)_{\zeta_1}=({\rm Im\,}\valp_j)_{\zeta_1'}$
and $({\rm Im\,}\valp_{j+1})_{\zeta_2}=({\rm Im\,}\valp_{j+1})_{\zeta_2'}$
for $(\zeta_1,\zeta_2)\neq (\zeta_1',\zeta_2')$ can be sometimes satisfied simultaneously.

Next we discuss the cancellations of the other factors $\Ups(\ups_{\zeta_1,\zeta_2})$ by some factors in the numerator,
as we did in \S\,\ref{sec:A2} and \S\,\ref{sec:A3}.
The ways of cancellations can be classified as table~\:\!\!\ref{tab:N},
as we listed in \S\,\ref{sec:3pt-func}.
\begin{table}[t]
\begin{center}
\begin{tabular}{|c||c|rl|}
\hline
case & denominator & \multicolumn{2}{|c|}{numerator} \\\hline\hline
$-$ & $\ups_{\zeta_1,\zeta_2}$ & \multicolumn{2}{|c|}{$-\!\!-$}\\\hline
1 & $\ups_{\zeta_1,\zeta_2+1}$ &
 $(Q\vrho-\valp_{j+1})\,\cdot$&\back\,$\ve_{\zeta_2}$\\\hline
2 & $\ups_{\zeta_1,\zeta_2^-}$ &
 $(Q\vrho-\valp_{j+1})\,\cdot$&\back\,$(\ve_{\zeta_2^-}+\cdots+\ve_{\zeta_2})$\\\hline
3 & $\ups_{\zeta_1^-,\zeta_2}$ &
 $(Q\vrho-\valp_j)\,\cdot$&\back\,$(\ve_{\zeta_1^-}\!+\cdots+\ve_{\zeta_1-1})$\\\hline
4 & $\ups_{\zeta_1^-,\zeta_2+1}$ &
 $(Q\vrho-(2Q\vrho-\valp_j))\,\cdot$&\back\,$(\ve_{\zeta_1^-}\!+\cdots+\ve_{\zeta_1-1})$\\\hline
\end{tabular}
\parcaption{Cancellations in a general quiver case \\[+1pt]
$(\zeta_1^-=1,\cdotsb,\zeta_1-1,~\zeta_2^-=1,\cdotsb,\zeta_2-1)$}
\label{tab:N}
\end{center}
\end{table}
Note that for the case 2 and 4, the condition $\valp_{j+1}\cdot \ve_{\zeta_2}=0$ is required, which is always satisfied.
Then after the cancellations for ${}^\forall (\zeta_1,\zeta_2)\in$ eq.\,(\ref{cond-zeta}), the remaining factors in the denominator are 
\ba
\text{denominator}:
&&\!\!\!
\Ups(\ups_{\zeta_1,\zeta_2}) \quad
\text{with}\quad
\zeta_1=1,\cdotsb,d_j\,,~
\zeta_2=1,\cdotsb,2d_j-d_{j-1}
\ea
Here we note that
\ba
(\zeta_1,\zeta_2) \text{~\,and\,~}
(\zeta_1-1,\zeta_2-\delta\zeta_2)\,\in\,\text{eq.\,(\ref{cond-zeta})}
\quad\Rightarrow\quad 
\delta\zeta_2=1 \text{~or~} 2\,.
\ea
If $\delta\zeta_2=1$, 
the case 3 for $(\zeta_1,\zeta_2)$ and the case 4 for $(\zeta_1-1,\zeta_2-1)$ means the cancellation of the same factors,
so they are never compatible.
Then in this case, we must give up the former cancellation, as we saw in the case of $SU(4)^{n-2}\times SU(3)\times SU(2)$ quiver. 
On the other hand, if $\delta\zeta_2=2$ or for $\min\,(\zeta_1,\zeta_2)\in$ eq.\,(\ref{cond-zeta}), the cancellations of all the case 1\,--\,4 can be done without any problem, as in the case of $SU(4)^{n-1}\times SU(2)$ quiver. 
There we gave up the case 2 cancellation for $(\zeta_1,\zeta_2)=(3,1)$,
but it is only because $\zeta_2^-$ does not run any value.

Therefore, after these cancellations, the remaining factors in the numerator are
\ba\label{numerator}
\text{numerator}:
&&\!\!\!\!
\begin{cases}
\,\Ups((Q\vrho-(2Q\vrho-\valp_j))\cdot\sum_{\zeta_j=\zeta_1^\flat}^{\zeta_1^\sharp}\ve_{\zeta_j}
&~\text{with}~~ \zeta_1^\flat,\zeta_1^\sharp=1,\cdots,d_j-1
\\[+5pt]
\,\Ups(Q\vrho-\valp_{j+1})\cdot\sum_{\zeta_{j+1}=\zeta_2^\flat}^{\zeta_2^\sharp} \ve_{\zeta_{j+1}}
&~\text{with}~~ \zeta_2^\flat,\zeta_2^\sharp=1,\cdots,d_{j+1}-1
\end{cases}\quad~
\ea
The latter case should be explained more.
The condition (\ref{cond-zeta}) means that 
the cancellation of case 1 and 2 leaves intact the following factors in the numerator other than eq.\,(\ref{numerator})\:\!:
$(Q\vrho-\valp_{j+1})\cdot (\ve_{\zeta^-}\!+\cdots+\ve_{\zeta})$
with
\ba\label{zeta-j}
\zeta~=~d_{j+1}\,,\cdotsb,\,2d_j-d_{j-1}\,,\,2d_j-d_{j-1}+2n\,,\,
d_{j-1}-d_{j-2}-1+i+\textstyle{\sum_{\ell=x}^{i-1}l_\ell}
\ea
where $n=1,\cdotsb,2d_{j-1}-d_j-d_{j-2}$ and $i=x,\cdotsb,s\,(=d_{j-1})$.
However, all these factors are removed by the case 3 cancellation in the next 3-point function $C(2Q\vrho-\valp_{j+1},\valp_{j+2};\vbet_{j+1})$.

Then the correspondence of parameters can be set as
\ba\label{mass-N3}
\text{$SU(d_j)$ adjoint scalar VEV}&&
\vha_j=i\valp_j'
\nn\\[+4pt]
\text{$SU(d_{j+1})$ adjoint scalar VEV}&&
\vha_{j+1}=i\valp_{j+1}'
\nn\\[+2pt]
\text{$SU(d_j)\times SU(d_{j+1})$ bifundamental mass}&& 
\nu_j=\frac{Q}{2}+im_j
\nt
\text{$SU(d_{j})$ fundamental mass}&& 
\mu_p^{(j)}=\frac{Q}{2}+im_j+i\alpha'_{j+1,p}
\ea
for $p=d_{j+1}+1,\cdotsb,2d_j-d_{j-1}$.
Note that the number of $SU(d_{j})$ fundamental matter fields is $2d_j-d_{j-1}-d_{j+1}\,(\geq 0)$, as we mentioned in eq.\,(\ref{quiver-cond}).

Finally, we can show the correspondence of 3-point function and the 1-loop partition function as 
\ba\label{corr-N3}
C(2Q\vrho-\valp_j,\valp_{j+1};\vbet_j)
\!&=&\!
A\,
f_{N-1}(m_j)h(2Q\vrho-\valp_j)h(\valp_{j+1})
\nt&&\times\,
\frac{1}{\Ups(0)^{N-d_j}}\,
\bigl|z^\tx{1lp}_\tx{bfd}(\vha_j,\vha_{j+1},\nu_j)\bigr|^2
\prod_{p=d_{j+1}+1}^{2d_j-d_{j-1}}\bigl|z^\tx{1lp}_\tx{fd}(\vha_j,\mu_p^{(j)})\bigr|^2
\nt&&\times\,
\prod_{\{\ve_{\zeta_j\!}\}}\Ups(i\valp_j\cdot \ve_{\zeta_j})
\prod_{\{\ve_{\zeta_{j+1}\!}\}}\Ups(-i\valp_{j+1}\cdot \ve_{\zeta_{j+1}})
\ea
with the conditions (\ref{cond-N}).
$\{\ve_{\zeta_j\!}\}$ and $\{\ve_{\zeta_{j+1}\!}\}$ are defined in eq.\,(\ref{numerator}).
Again, as in eq.\,(\ref{vec}), 
the last two factors correspond to the factors 
$z^\tx{1lp}_\tx{vec}(\vha_j)$ and $z^\tx{1lp}_\tx{vec}(\vha_{j+1})$,
together with the factors from the next 3-point functions.

\sssn{Type 4\;\!: the last 3-point function $C(2Q\vrho-\valp_n,\vbet_0;\vbet_n)$}

The momenta of Toda vertex operators are set as
\ba\label{set-N4}
&&\back
\valp_n=Q\vrho_{Y_n}+i\,\Bigl[(\valp'_n,\vec 0)+\vgam_{j\:\![d_n,l_{d_n+1},\cdotsb,l_s]}\Bigr],\quad
\vbet_n=\left(\frac{Q}{2}+im_n\right)N\vome_{N-1}\,,\nt
&&\back
\vbet_0=Q\vrho_{Y}+i\vbet'_{0\,Y}\,,
\ea
This is just a reparametrization of eq.\,(\ref{set-N3}),
so the discussion is almost parallel to type 3.
Then if the correspondence of parameters is set as
\ba\label{mass-N4}
\text{$SU(d_n)$ adjoint scalar VEV}&&
\vha_n=i\valp_n'
\nn\\[+2pt]
\text{$SU(d_n)$ fundamental mass}&& 
\mu_p=\frac{Q}{2}+im_n+i\beta'_{0,p}
\ea
for $p=1,\cdotsb,2d_n-d_{n-1}$,
we can show the correspondence of 3-point function and the 1-loop partition function as 
\ba\label{corr-N4}
C(2Q\vrho-\valp_n,\vbet_0;\vbet_n)
\!&=&\!
A\,
g(\vbet_0)f_{N-1}(m_n)h(2Q\vrho-\valp_n)
\nt&&\times\,
\frac{1}{\Ups(0)^{N-d_n}}
\prod_{p=1}^{2d_n-d_{n-1}}\bigl|z^\tx{1lp}_\tx{fd}(\vha_n,\mu_p)\bigr|^2
\prod_{\{\ve_{\zeta_n}\}}\Ups(i\valp'_n\cdot \ve_{\zeta_n})
\ea
with the conditions
\ba\label{cond-N2}
m_n=({\rm Im\,}\valp_n)_{\zeta_1}-({\rm Im\,}\vbet_0)_{\zeta_2}
\qquad\text{for}~~
(\zeta_1,\zeta_2)\in \text{eq.\,(\ref{cond-zeta})\, with}~\,j=n\,.
\ea
As we defined in eq.\,(\ref{prod-e}),
when we take the product $\prod'_{e>0}$ in $g(\vbet_0)$,
the following factors must be removed:
$\Ups\bigl((Q\vrho-\vbet_0)\cdot(\ve_{\zeta^-}\!+\cdots+\ve_{\zeta})\bigr)$
for ${}^\forall\zeta\in\{\zeta_2~\text{in eq.\,(\ref{cond-zeta}) with}~j=n\}$.

\sssn{Summary}

By putting all the results together,
{\em i.e.}~eq.\,(\ref{corr-N1}), (\ref{corr-N2}), (\ref{corr-N3}) and (\ref{corr-N4}),
we can show that 
the whole of Toda correlation function with descendant level 0 (\ref{tree}) exactly corresponds to the 1-loop partition function of gauge theory with a general quiver gauge group (\ref{quiver}) as
\ba
V_\emp\,=\,
A^{n+1}{\:\!}h(2Q\vrho)^n{\:\!}g(\vbet_\infty){\:\!}g(\vbet_0)
\prod_{k=1}^{n+1}f_{N-1}(m_k)
\prod_{j=1}^n \frac{1}{\Upsilon(0)^{N-d_j}}
\prod_{p<q}{\:\!}(\alpha'_{j,p}-\alpha'_{j,q})^2{\:\!}|Z_\tx{1-loop}|^2~~
\ea
if we set the correspondence of parameters as eq.\,(\ref{mass-N1}), (\ref{mass-N2}), (\ref{mass-N3}) and (\ref{mass-N4}),
and impose the conditions (\ref{cond-N}) and (\ref{cond-N2}).
Each pole $1/\Ups(0)$ corresponds to one of these conditions, 
{\em i.e.}~the setting of mass of a hypermultiplet. The total order of poles is $\sum_{j=1}^n (N-d_j)$, which is equal to the number of times which Hanany-Witten transition occurs in the D4/NS5/D6-brane system, 
as we discussed in \S\,\ref{sec:3pt-func}.
Therefore, we believe that we can properly understand the physical interpretation of all the poles in 3-point correlation functions of $A_{N-1}$ Toda theory in AGT-W relation.

\section{Conclusion and Discussion}

In this paper, we show the correspondence between the correlation function of $A_{N-1}$ Toda theory with descendant level 0 and the 1-loop part of partition function of $\cN=2$ $SU(N)$ quiver gauge theory with a general quiver gauge group.
All the parameters except gauge coupling constants appear in this part,
so in this sense, we claim that the ansatz for correspondence of parameters in AGT-W relation~\cite{Kanno:2009,Drukker:2010} is completely justified.

The remaining part of AGT-W relation is the correspondence between the descendant part of correlation function of Toda theory and the instanton part of partition function of gauge theory.
Now the correspondence of parameters is clearly understood, 
then the only unclear point is so-called `$U(1)$ factor' in AGT-W relation~\cite{Alday:2009}.
At this moment, there seems to be no consensus among researchers with regard to the way of determining this factor,
while some researchers propose that this factor is nothing but the free string amplitude~\cite{Alba:2009}.


On this problem, we have a direction of discussion.
In our previous paper~\cite{Kanno:2011}, we pointed out that
$\cW_{1+\infty}$ algebra may exist as a symmetry behind AGT-W relation,
by showing that Toda correlation function plus the $U(1)$ factor can be simply written in terms of this algebra.
In this algebra, $U(1)$ generator naturally coexists with $\cW_N$ generators of $A_{N-1}$ Toda theory.
Therefore, we consider that this $U(1)$ generator in $\cW_{1+\infty}$ algebra must be related to the $U(1)$ factor in AGT-W relation.
Then from this viewpoint, it may be possible to justify the interpretation of the $U(1)$ factor as the free string amplitude.

Anyway in order to check the remaining part of AGT-W relation, we must calculate the correlation function of Toda theory with an arbitrary descendant level.
The most basic way is to calculate the inverse Shapovalov matrix of each level as it was done in \cite{Alba:2009,Mironov:2009,Kanno:2010}, 
but it must be not a realistic way if we want to calculate in an arbitrary high level.
One choice is the calculation by Dotsenko-Fateev method, 
which has been recently discussed also in the context of AGT-W relation~\cite{Mironov:2010zs,Mironov:2010,Zhang:2011}.
Up to now, however, these discussions are restricted to the case of
4-point correlation function and $Q=b+b^{-1}=0$.
Especially, it must be very difficult to discuss the case of $Q\neq 0$,
so all we can do may be to calculate the correlation function with arbitrary number of points and descendant level, but $Q=0$.

Finally, we want to say that AGT-W relation is still a very strange relation.
In particular, we discuss the considerably general cases of $\cN=2$ $SU(N)$ quiver gauge theory, but the corresponding correlation function of Toda theory seems in very special cases. 
Through the further various investigations, we hope to understand what it means from the viewpoint of, for example, 
$\cW_{1+\infty}$ algebra, superconformal theory, and M5-brane dynamics.

\subsection*{Acknowledgments}
We would like to thank Yutaka Matsuo and Shoichi Kanno for useful discussions and comments.
The author is partially supported by Grant-in-Aid ({\#}23-7749) for JSPS fellows.


\appendix

\section{Partition function of $\cN=2$ $SU(N)$ quiver gauge theory}
\label{sec:Nek}

The full partition function of 4-dim $\cN=2$ $SU(N)$ quiver gauge theory can be written as
\ba
Z=Z_\tx{class}\,Z_\tx{1-loop}\,Z_\tx{inst}\,.
\ea
We see each part of function in the following~\cite{Nekrasov:2002,Nekrasov:2003}.

\paragraph{Classical part}

The classical part of the partition function is
\ba
Z_\tx{class}=\exp\left[\sum_{k=1}^n 2\pi i \tau_k |\vec{\hat a}_k|^2\right]
\ea
where $\tau_k:=\frac{\theta_k}{2\pi}+\frac{4\pi i}{g_k^2}$ is
the complex UV coupling constant, and
$\vec{\hat a}_k:=\sum_{i=1}^{d_k-1}a_i\vec{e}_i$ is
the diagonal of VEV's $a_i$ of adjoint scalars.
$\vec e_i$ are the simple roots of gauge symmetry algebra,
which are usually defined as eq.\,(\ref{root}) for $SU(N)$ algebra.
It gives, for example,
$\vec{\hat a}=(a_1,-a_1)$ for $SU(2)$ and
$\vec{\hat a}=(a_1,-a_1+a_2,-a_2)$ for $SU(3)$.

\paragraph{1-loop part}
The 1-loop contribution to the partition function
is
\ba\label{Z1lp}
Z_\text{1-loop}
&=&
 \left(\prod_{k=1}^n z^\tx{1lp}_\tx{vec}(\vha_k)\right)
 \left(\prod_{\brp=1}^{d_1} z^\tx{1lp}_\tx{afd}(\vha_1,\bmu_\brp)\right)
\nt&&\times
 \left(\prod_{k=1}^{n-1} z^\tx{1lp}_\tx{bfd}(\vha_k,\vha_{k+1},m_k)\right)
 \left(\prod_{p=1}^{d_n} z^\tx{1lp}_\tx{fd}(\vha_n,\mu_p)\right)
\ea
where
$\mu_p$, $\bar\mu_\brp$, $m_k$ are the mass of fundamental, antifundamental and bifundamental fields, respectively.
The functions $z^\tx{1lp}$ are defined as
\ba\label{1lpfactor}
z^\tx{1lp}_\tx{vec}(\va)
&=&
\prod_{i<j}\exp\left[-\gamma_{\eps_1,\eps_2}(\ha_i-\ha_j-\eps_1)
-\gamma_{\eps_1,\eps_2}(\ha_i-\ha_j-\eps_2)\right],\nt
z^\tx{1lp}_\tx{fd}(\va,\mu)
&=&
\prod_{i}\exp\left[\gamma_{\eps_1,\eps_2}(\ha_i-\mu)\right],\nt
z^\tx{1lp}_\tx{afd}(\va,\bar\mu)
&=&
\prod_{i}\exp\left[\gamma_{\eps_1,\eps_2}(-\ha_i+\bar\mu-\eps_+)\right],\nt
z^\tx{1lp}_\tx{bfd}(\va,\vb,m)
&=&
\prod_{i,j}\exp\left[\gamma_{\eps_1,\eps_2}(\ha_i-\hb_j-m)\right],
\ea
where
$\eps_+:=\eps_1+\eps_2$
($\eps_1$, $\eps_2$ are Nekrasov's deformation parameters).
The function $\gamma_{\eps_1,\eps_2}(x)$ is related to double Gamma
function $\Gamma_2(x|\eps_1,\eps_2)$ as
\ba
\gamma_{\eps_1,\eps_2}(x)=\log \Gamma_2(x+\eps_+|\eps_1,\eps_2)\,,
\ea
and the double Gamma function is defined with the double zeta function as
\ba
\Gamma_2(x|\eps_1,\eps_2):=\exp\left[
\frac{d}{ds}\bigg|_{s=0}\zeta_2(s;x|\eps_1,\eps_2)
\right]\,,
\ea
and the double zeta function is defined as
\ba
\zeta_2(s;x|\eps_1,\eps_2)
=\sum_{m,n}(m\eps_1+n\eps_2+x)^{-s}
=\frac{1}{\Gamma(s)}\int_0^\infty
\frac{dt}{t}
\frac{t^s e^{-tx}}{(1-e^{-\eps_1t})(1-e^{-\eps_2t})}\,.
\ea
When we discuss AGT-W relation, we often use the properties of double Gamma function
\ba\label{prop-gamma}
\Gamma_2(x^*)=\Gamma_2(x)^*\,,\quad
\Gamma_2(x+\eps_1)\Gamma_2(x+\eps_2)
=x\Gamma_2(x)\Gamma_2(x+\eps_+)\,,
\ea
and the relation to Upsilon function
\ba
\Upsilon(x)=\frac{1}{\Gamma_2(x|b,b^{-1})\Gamma_2(Q-x|b,b^{-1})}\,.
\ea

\paragraph{Instanton part}

The instanton contribution is obtained by Nekrasov's instanton counting
formula with Young tableaux as
\ba
Z_\tx{inst}
&=&\sum_{\{\vY_1,\cdotsb,\vY_n\}}
 \left(\prod_{k=1}^n q_k^{|\vY_k|}z_\tx{vec}(\vha_k,\vY_k)\right)
 \left(\prod_{\brp=1}^{d_1} z_\tx{afd}(\vha_1,\vY_1,\bmu_\brp)\right)
\nt&&\times
 \left(\prod_{k=1}^{n-1} z_\tx{bfd}(\vha_k,\vY_k;\vha_{k+1},\vY_{k+1};m_k)\right)
 \left(\prod_{p=1}^{d_n} z_\tx{fd}(\vha_n,\vY_n,\mu_p)\right)
\ea
where
$q_k:=e^{2\pi i\tau_k}$ ($\tau_k$ is the coupling constant),
and
$\vec Y_k=(Y_{k,1},\cdotsb,Y_{k,d_k})$ is a set of Young tableaux.
$|\vec Y_k|$ is the total sum of number of boxes of Young tableaux
$Y_{k,i}$ ($i=1,\cdotsb,d_k$).
Each factor of the instanton part is written as
\ba
z_\tx{bfd}(\vha,\vY;\vhb,\vW;m)&=&
 \prod_{i,j} \prod_{s\in Y_i}(E(\ha_i-\hb_j,Y_i,W_j,s)-m)
\nt&&\times
 \prod_{t\in W_j}(\eps_+-E(\hb_j-\ha_i,W_j,Y_i,t)-m)\,,\nt
z_\tx{vec}(\vha,\vY)&=&1/z_\tx{bfd}(\vha,\vY;\vha,\vY;0)\,,\nt
z_\tx{fd}(\vha,\vY,\mu)&=&
 \prod_{i} \prod_{s\in Y_i}(\phi(\ha_i,s)-\mu+\eps_+)\,,\nt
z_\tx{afd}(\vha,\vY,\bmu)&=&z_\tx{fd}(\vha,\vY,\eps_+-\bmu)\,.
\ea
The functions $E(\ha,Y,W,s)$ and $\phi(\ha,s)$ are defined as
\ba
E(\ha,Y,W,s)
&=&\ha-\eps_1(\lambda'_{W,j}-i)+\eps_2(\lambda_{Y,i}-j+1)\,,\nt
\phi(\ha,s)&=&\ha+\eps_1(i-1)+\eps_2(j-1)\,,
\ea
where $s=(i,j)$ denotes the position of the box in a Young tableau
({\em i.e.}~the box in $i$-th column and $j$-th row).
$\lambda_{Y,i}$ is the height of $i$-th column, and
$\lambda'_{Y,j}$ is the length of $j$-th row for Young tableau $Y$.
That is,
$\lambda'_{Y,j}-i$ 
and
$\lambda_{Y,i}-j$ 
are the length of `leg' and `arm' of the Young tableau $Y$
for the box $s=(i,j)$, respectively.

\end{document}